\documentclass[twocolumn]{autart}    % Enable this line and disable the 

\newif\ifappendix
\global\appendixtrue

\usepackage{graphicx} % for pdf, bitmapped graphics files
\usepackage{enumitem}
\usepackage{amsmath} % assumes amsmath package installed
\usepackage{amssymb}  % assumes amsmath package installed
\usepackage{mathtools}
\usepackage{xcolor}

\usepackage{subcaption}

\usepackage[noend]{algpseudocode}

\usepackage{bbm}

\newcommand{\defeq}{\vcentcolon=}

\DeclareMathOperator*{\argmax}{arg\,max}
\DeclareMathOperator*{\argmin}{arg\,min}

\newtheorem{property}{Property}

\let\emptyset\varnothing

\newcommand\blfootnote[1]{%
  \begingroup
  \renewcommand\thefootnote{}\footnote{#1}%
  \addtocounter{footnote}{-1}%
  \endgroup
}

\usepackage{tikz}
\usetikzlibrary{shapes.geometric, arrows.meta, positioning}
\tikzstyle{startstop} = [rectangle, rounded corners, minimum width=1cm, minimum height=2cm,text centered, draw=black, fill=pink, text width=0.5cm]
\tikzstyle{IO} = [ellipse, minimum width=2.5cm, minimum height=0.1cm,text centered, draw=black, text width=3.5cm]
\tikzstyle{arrow} = [thick,->,>={Stealth}]
\tikzset{font={\fontsize{10pt}{12}\selectfont}}
\usetikzlibrary{patterns}
\usepackage{pgfplots}
\tikzstyle{bag} = [align=left]
\usetikzlibrary{shapes.misc}

\usetikzlibrary{decorations.pathmorphing}
\tikzset{zigzag/.style={decorate,decoration=zigzag}}
\definecolor{sandybrown}{rgb}{0.96, 0.64, 0.38}

\usepackage{url}

\begin{document}

\begin{frontmatter}

\title{Data-Driven Certificate Synthesis}

\author[Oxford]{Luke Rickard}\ead{rickard@robots.ox.ac.uk}
\author[OxfordCS]{Alessandro Abate}
\author[Oxford]{Kostas Margellos}
\address[Oxford]{Department of Engineering Science, University of Oxford}
\address[OxfordCS]{Department of Computer Science, University of Oxford}


\begin{abstract}
    We investigate the problem of verifying different properties of discrete time dynamical systems, namely, reachability, safety and reach-while-avoid. 
    To achieve this, we adopt a data driven perspective and, using past system trajectories as data, we aim at learning a specific function termed \emph{certificate} for each property we wish to verify. 
    % The certificate construction problem is treated as a safety informed neural network training process, where we use a neural network to learn the parameterization of each certificate, while the
    We seek to minimize a loss function, designed to encompass conditions on the certificate to be learned that encode the satisfaction of the associated property. 
    Besides learning a certificate, we quantify probabilistically its generalization properties, namely, how likely it is for a certificate to be valid (and hence for the associated property to be satisfied) when it comes to a new system trajectory not included in the training data set.
    We view this problem under the realm of probably approximately correct (PAC) learning under the notion of compression, and use recent advancements of the so-called scenario approach to obtain scalable generalization bounds on the learned certificates. 
    To achieve this, we design a novel algorithm that minimizes the loss function and hence constructs a certificate, and at the same time determines a quantity termed compression, which is instrumental in obtaining meaningful probabilistic guarantees. 
    This process is novel per se and provides a constructive mechanism for compression set calculation, thus opening the road for its use to more general non-convex optimization problems. 
    We verify the efficacy of our methodology on several numerical case studies, and compare it (both theoretically and numerically) with closely related results on data-driven property verification.
\end{abstract}

\end{frontmatter}

\section{Introduction}
{\blfootnote{This work was supported by the EPSRC Centre for Doctoral Training in Autonomous Intelligent Machines and Systems EP/S024050/1.}}

%\subsection{Motivation}
Dynamical systems offer a rich class of models for describing the behavior of diverse, complex systems \cite{Hirsch2003DifferentialED}. 
It is often of importance that these systems meet certain properties, for example, stability, safety or reachability~\cite{DBLP:conf/hybrid/EdwardsPA24,DBLP:journals/tac/JagtapSZ21,DBLP:journals/tac/NejatiLJSZ23,DBLP:journals/tac/PrajnaJP07}.
Verifying the satisfaction of these properties, also termed as specifications, is a challenging, but important, problem.

One research direction involves discretizing the state space \cite{Abate2025,DBLP:journals/jair/BadingsRAPPSJ23,DBLP:conf/qest/RickardBRA23}, in order to construct a finite model with guarantees generated using probabilistic or statistical model checkers \cite{DBLP:journals/sttt/BadingsCJJKT22,DBLP:conf/l4dc/RickardAM24}.
Discretizing the state space however tends to be computationally expensive, even for low-dimensional systems.
% Such techniques frequently suffer from the so-called ``curse of dimensionality'' which prohibits their application to systems of higher dimension.}

\begin{table*}[t]
    \centering
        \caption{\begin{center} Classification of Certificate Synthesis Approaches \end{center}  $^\star$~Theorem~\ref{thm:Guarantees}, and Algorithms \ref{algo:sub} \& \ref{algo:main}, follow a non-convex scenario approach methodology, that does not require knowledge of the Lipschitz constant of the dynamics, and offer probabilistic verification bounds that do not necessarily scale exponentially in the state space dimension as with \cite{DBLP:journals/tac/NejatiLJSZ23,DBLP:journals/automatica/SalamatiLSZ24}.}
    \label{tab:litt}
    \begin{tabular}{p{5cm}|p{5.5cm}|p{6cm}}
         % Model-Based Approaches& \multicolumn{2}{|c}{Data-Driven Approaches}
         % \\ \cline{2-3}
         Model-Based Synthesis \& Guarantees & Data-Driven Synthesis \& Model-Based Guarantees %Data-Driven Optimization 
         & Data-Driven Synthesis \& Guarantees \\ \hline
         %Guarantees Scale Exponentially in State Space Dimension & Guarantees Independent of State Space Dimension \\ \hline
         Sum of Squares Programming~\cite{DBLP:conf/cdc/Papachristodoulou02}& Counter-Example Guided Inductive Synthesis \cite{DBLP:conf/hybrid/AbateAEGP21,DBLP:conf/nips/ChangRG19,DBLP:conf/cdc/DaiLPT20,DBLP:conf/hybrid/EdwardsPA24} & Neural Network Techniques~\cite{ANAND20232431,DBLP:conf/corl/SunJF20} \\ 
         MPC + Reachability analysis~\cite{10.1007/978-3-031-65633-0_18,DBLP:conf/ijcai/RenLL0023} & Neural Hamilton-Jacobi Reachability Analysis~\cite{solanki2025certifiedapproximatereachabilitycare,DBLP:journals/corr/abs-2407-20532} & Convex Scenario Optimization~\cite{DBLP:journals/tac/NejatiLJSZ23,DBLP:journals/automatica/SalamatiLSZ24} \\
         SAT-modulo-theory synthesis\cite{bcAPA20} & Neural Certificates for Safety~\cite{DBLP:journals/corr/abs-2006-08465}  
         &Theorem~\ref{thm:Guarantees}, Algorithms \ref{algo:sub} \& \ref{algo:main}$^\star$\\    
         % & \\  & & \\
         %  & \\ 
    \end{tabular}
\end{table*}

%\subsection{Certificates}
An alternative approach to verify properties of dynamical systems that does not require discretizing the state space, is through the use of \emph{certificates} \cite{Abate2025,DBLP:conf/eucc/AmesCENST19,DBLP:conf/hybrid/PrajnaJ04}.
The goal is to determine a function over the system's state space that exhibits certain properties.
A well-investigated example of such certificate is that of a Lyapunov function, used to verify that dynamics satisfy some stability property \cite{Lyapunov1994TheGP}. Here
we consider constructing reachability, safety, and reach-while-avoid (RWA) certificates for discrete-time systems. 
Overviews of techniques for certificate learning can be found in~\cite{Abate2025,DBLP:journals/trob/DawsonGF23}; Table~\ref{tab:litt} summarizes the related literature, which we discuss below.
% In general, constructing such certificates is a case dependent task and requires domain specific expertise. 
% Restricting the class of certificates to polynomial functions, 
% they can be obtained by solving a convex sum-of-squares problem \cite{DBLP:conf/cdc/Papachristodoulou02}.

% The construction of such certificates has garnered significant research activity in recent years. 
One approach to certificate synthesis considers verifying the behavior of systems assuming that a model of the underlying dynamical system is known. 
Restricting the class of models to polynomial functions, certificates can be obtained by solving a convex sum-of-squares problem~\cite{DBLP:conf/cdc/Papachristodoulou02}. More generally, synthesis approaches leveraging SAT-modulo-theories can alternatively be leveraged~\cite{bcAPA20,DBLP:journals/trob/DawsonGF23}.  
Availability of a model allows for co-design of a controller meeting certain specifications, such as safety and reachability~\cite{10.1007/978-3-031-65633-0_18,DBLP:conf/ijcai/RenLL0023}, using tools based on Model Predictive Control (MPC), or reachability analysis.

The inclusion of the model's knowledge in the synthesis procedure in practice limits  the complexity of the models that may be studied.
% Such classical approaches are inherently based on the knowledge and study of model's dynamics. 
To alleviate this requirement, data-driven techniques, such as counter-example guided inductive synthesis (CEGIS)~\cite{DBLP:conf/hybrid/AbateAEGP21,DBLP:conf/nips/ChangRG19,DBLP:conf/cdc/DaiLPT20,DBLP:conf/hybrid/EdwardsPA24} are able to synthesize certificates for general non-linear systems. 
This is achieved via the use of neural networks as certificate templates, allowing for the approximation of any function within a certain function space~\cite{DBLP:journals/nn/HornikSW89}. 
Neural networks have also been explored as a tool in~\cite{solanki2025certifiedapproximatereachabilitycare,DBLP:journals/corr/abs-2407-20532} for guaranteeing reachability, and in~\cite{DBLP:journals/corr/abs-2006-08465} for safety analysis. 

Such approaches involve data-driven synthesis, however, they still require a model of the system when it comes to providing guarantees on the synthesized certificates.
Obtaining a model of the system is in general difficult, as it requires domain-specific knowledge. 
To alleviate these issues, in this work we follow a data-driven route that is model-free as far both the certificate synthesis and guarantee process is concerned. 
One way to achieve this involves using one set of samples to synthesize a certificate, and a separate set for validation~\cite{DBLP:conf/corl/SunJF20}. %\textcolor{red}{[placing of sentence before is possibly out of place?]}
An alternative approach which does not require different samples for validation and is hence also the one most closely related to our formulation, involves probabilistic property satisfaction using a convex design and results on scenario optimization~\cite{DBLP:journals/tac/NejatiLJSZ23,DBLP:journals/automatica/SalamatiLSZ24}, and extensions to neural networks~\cite{ANAND20232431,DBLP:conf/corl/SunJF20}. 
However, these developments rely on the system dynamics being Lipschitz continuous and for the Lipschitz constant to be known (or a bound on this to be available). 
Moreover, the probabilistic guarantees provided exhibit an exponential growth with respect to the system dimension. Both issues are not present within our proposed approach.
% Unfortunately, existing convex and neural network approaches both suffer from guarantees which scale exponentialy in the state space dimension.

In this work, we follow a scenario approach paradigm as in~\cite{DBLP:journals/tac/NejatiLJSZ23,DBLP:journals/automatica/SalamatiLSZ24}, however, we exploit some different statistical learning theoretic developments in scenario optimization. This allows us to remove the requirements for 
convexity and knowledge of the Lipschitz constant of the dynamics, and establish probabilistic verification bounds that do not necessarily scale exponentially on the state dimension, but their complexity rather depends on the complexity of the underlying property verification task. 

In particular, we
use any parameterized function approximator as certificate template, and learn these parameters using a finite number of system trajectories treated as samples.
% In this work, we develop a data-driven technique that can make use of any parameterized function approximator as certificate template, and that provides probably approxmately correct (PAC) guarantees that scale independently of the state space dimension.
We formulate the certificate synthesis problem as a (possibly) non-convex optimization
program, that involves minimizing an appropriately designed loss function, whose minimum value implies that a given property is satisfied. To minimize that loss function we also 
design a subgradient descent style procedure~\cite{DBLP:books/cu/BV2014}. We accompany the synthesized certificate with \textit{probably approximately correct} (PAC) guarantees on its validity, and hence on the probability of of satisfying the underlying property, when it comes to a new system trajectory. It is to be noted that such a procedure does not require using a separate data-set for validation. 
To establish such PAC guarantees, we make use of bounds on the change of a quantity termed \textit{compression set} (namely, a subset of the data which would return the same result as the entire set)~\cite{DBLP:journals/jmlr/CampiG23,Floyd_Warmuth_1995,DBLP:journals/tac/MargellosPL15}, through recent advancements of the so-called \textit{scenario approach} \cite{Scen_approach_book,DBLP:journals/siamjo/CampiG08,DBLP:journals/mp/CampiG18,DBLP:journals/tac/CampiGR18,DBLP:journals/mp/GarattiC22}. 
In particular, we are inspired by the novel theoretical \textit{pick-to-learn framework}~\cite{DBLP:conf/nips/PaccagnanCG23}, which provides a meta-algorithm for calculating a compression set with favourable properties. 
Here we extend the scope of the pick-to-learn framework by providing a constructive instance of the general framework to compute the cardinality of compression sets for non-convex optimization.

Our main contributions can be summarized as follows: 
\begin{enumerate}
    \item We develop a novel methodology for the synthesis of certificates to verify a wide class of properties, namely, reachability, safety and reach-while-avoid specifications, of discrete-time dynamical systems. 
    Our results complement the ones in \cite{DBLP:journals/sttt/BadingsCJJKT22} which are concerned with direct property verification and do not construct certificates. 
    Our framework constitutes a first step towards control synthesis exploiting the constructed certificates.
    \item 
    Capitalizing on developments on scenario optimization using the notion of compression, we accompany the constructed certificates with probabilistic guarantees on their generalization properties, namely, on how likely it is that the certificate remains valid when it comes to a new system trajectory. 
    We contrast our approach with \cite{DBLP:journals/tac/NejatiLJSZ23} and discuss the relative merits of each, both theoretically (Section \ref{sec:related}) and numerically (Section \ref{sec:exp}).
    \item As a byproduct of our certificate construction algorithm, we provide a novel mechanism to compute the \emph{compression set}, which is instrumental in obtaining meaningful probabilistic guarantees. 
    This results in \emph{a posteriori} bounds which, however, scale favorably with respect to the system dimension. 
    This process is novel per se and provides a constructive approach for the general compression set calculation in \cite{DBLP:conf/nips/PaccagnanCG23}, opening the road for its use in general non-convex optimization problems.
\end{enumerate}
%\section{Notation}
% \\

\emph{Notation.}
We use $\{\xi_k\}_{k=0}^K$ to denote a sequence indexed by $k \in \{0,1,\dots,K \}$. $V \models \psi$ defines condition satisfaction i.e., it evaluates to true if the quantity $V$ on the left satisfies the condition $\psi$ on the right, e.g., $x = 1 \models x > 0$ evaluates to true and $x=-1 \models x > 0$ evaluates to false.  Using $\not\models$ represents the logical inverse of this (i.e., condition dissatisfaction). By $(\forall \xi \in \Xi) V\models\psi(\xi)$ we mean that some quantity $V$ satisfies a condition $\psi$ which, in turn, depends on some parameter $\xi$, for all $\xi \in \Xi$.

\section{Certificates}
\label{sec:certs}

We consider a family of certificates that allow us to make statements on the behavior of a dynamical system. %, namely, how likely it is that it satisfies certain properties.
Hence, we begin by defining a dynamical system, before considering the certificates and properties they verify.

\subsection{Discrete-Time Dynamical Systems}

We consider a bounded state space $X \subset \mathbb{R}^n$, and a dynamical system whose evolution starts at an initial state $x(0) \in X_I$, where $X_I\subseteq X$ denotes the set of all possible initial conditions. 
From an initial state, we can uncover a finite trajectory, i.e., a sequence of states $\xi = \{x(k)\}_{k=0}^T$, where $T\in \mathbb{N}_+$, by following the dynamics
\begin{equation}
\label{eq:Dyn}
    x(k+1)=f(x(k)).
\end{equation}
We define $f \colon X \rightarrow \mathbb{R}^n$, and assume it to permit unique solutions, but make no further assumptions on its properties.
The set of all possible trajectories $\Xi \subseteq X_I \times X^{T}$ is then the set of all trajectories starting from the initial set $X_I$.
This set-up considers only deterministic systems, but our methods are applicable to systems with stochastic dynamics - we discuss this in further detail in Section \ref{sec:learn_certs:data}. 
% We do not consider controller synthesis here, but recognize that o
Our general form of dynamical system allows for verifying systems with controllers ``in the loop'': for instance, our techniques allow us to verify the behavior of a system with a predifined control law structure, such as Model Predictive Control~\cite{DBLP:journals/automatica/GarciaPM89}.

In Section~\ref{sec:learn_certs}, we discuss using a finite set of trajectories in order to provide generalization guarantees for future trajectories. 
Our techniques only require a finite number of samples, and are \emph{theoretically} not restricted on the properties of such samples (for instance, we may have a finite number of samples each with an infinitely long time horizon). 
However, we discuss in Section~\ref{sec:training} how one can synthesize a certificate in practice, and our algorithms are required to store, and perform some calculations on, these trajectories (which is not \emph{practically} possible for $T$ taken to infinity, or continuous time trajectories).

% We are interested in verifying whether the behavior of a dynamical system satisfies certain properties.
% We use $\phi(\xi)$ to refer to a property of interest (defined concretely in the sequel), which is evaluated on a trajectory $\xi \in \Xi$. 
% Specifically, we will define conditions $\psi^s$, that will have to be satisfied over some sub-domains of the state space, and $\psi^\Delta (\xi)$ that will define conditions that will have to be satisfied only at specific points along a trajectory $\xi \in \Xi$. The separate notations $\psi^s$ and $\psi^\Delta$ are used to distinguish between trajectory-independent and -dependent conditions, respectively.
    
In order to verify the satisfaction of a property $\phi$, we consider the problem of finding a \emph{certificate} as follows.

\begin{defn}[Property Verification \& Certificates]\label{prob:property_cert}
    Given a property $\phi(\xi)$, and a function $V\colon \mathbb{R}^n \rightarrow \mathbb{R}$, let $\psi^s$ and $\psi^\Delta (\xi)$ be conditions such that, if$$
            \left[\exists V \colon \left( V \models \psi^s\wedge (\forall \xi \in \Xi) V\models\psi^\Delta(\xi)\right)\right] \implies \phi(\xi), \forall \xi \in \Xi,$$
    then the property $\phi$ is verified for all $\xi \in \Xi$. We then say that such a function $V$ is a \emph{certificate} for the property encoded by $\phi$. 
\end{defn}
     
In words, the implication of Definition \ref{prob:property_cert} is that if a certificate $V$ satisfies the trajectory-independent conditions in $\psi^s$, as well as the trajectory-dependent conditions in $\psi^\Delta (\xi)$, for all $\xi \in \Xi$, then the property $\phi(\xi)$ is satisfied for all trajectories $\xi \in \Xi$. 
\subsection{Certificates}

We now provide a concrete definition for a number of these properties, and associated certificates (and certificate conditions) that meet the format of Definition \ref{prob:property_cert}. 
% We fix a time horizon $T<\infty$. 
We assume that $V$ is continuous, so that when considering the supremum/infimum of $V$ over a bounded set, this is well-defined. %$X$ (already assumed to be bounded) or over some of its subsets, this is well-defined.

\begin{property}[Reachability]\label{prop:reach}
Consider \eqref{eq:Dyn}, and let
    $X_G, X_I \subset X$ denote a goal and initial set, respectively. Assume further that $X_G$ is compact and $\partial X_G$ denotes its boundary. If, for all $\xi \in \Xi$, 
    \begin{equation}
        \phi_{\mathrm{reach}}(\xi) \defeq \exists k \in \{0,\dots,T\} \colon x(k) \in X_G, 
        % &\wedge~ \forall j \in \{0,\dots,k\} \colon x(j) \in X
    \end{equation}
    %     \begin{align}
    %     \phi_{\mathrm{reach}}(\xi) \defeq &\exists k \in \{0,\dots,T\} \colon x(k) \in X_G, \nonumber \\
    %     &\wedge~ \forall j \in \{0,\dots,k\} \colon x(j) \in X
    % \end{align}
    holds, then we say that $\phi_{\mathrm{reach}}$ encodes a reachability property.
    $\Xi$ denotes the set of trajectories consistent with \eqref{eq:Dyn} and with initial states contained within $X_I$.
\end{property}
By the definition of $\phi_{\mathrm{reach}}$ it follows that verifying that a system exhibits the reachability property is equivalent to verifying that all trajectories generated from the initial set enter the goal within at most $T$ time steps. %, and stay within the domain $X$ till then.
To verify this property, we consider a certificate that must satisfy a number of conditions. 
These conditions are summarized next. Fix $\delta> -\inf_{x \in X_I} V(x) \geq 0$.  We then have
\begin{align}
	\label{eq:reach_init}	
  &V(x) \leq 0, \; \forall x \in X_I,\\
 \label{eq:reach_goal}
		&V(x) \geq -\delta, \; \forall x \in \partial X_G, \\
  \label{eq:reach_else}
		&V(x) > -\delta, \; \forall x \in X \setminus X_G,\\
        &V(x) > 0, \; \forall x \in \mathbb{R}^N \setminus X,
  \label{eq:reach_dom_border}\\
		&V(x(k+1)) - V(x(k)) \label{eq:reach_deriv} \\ 
        & <- \frac{1}{T} \Big( \sup_{x \in X_I} V(x) + \delta \Big),~ k=0,\dots,k_G-1, \nonumber %\\&\;\forall \{x(k)\}_{k=0,\dots,T} \colon x(0) \in X_I,
		% & \qquad \qquad \forall x(k) \in X \setminus X_G,
\end{align}
       where $k_G \defeq \min \{k \in \{0,\dots,T\} \colon V(x(k)) \leq -\delta\}$, or $k_G=T$, if there is no such $k$.
Conditions (\ref{eq:reach_goal})-(\ref{eq:reach_dom_border}) allow characterizing different parts of the state space by means of specific level sets of $V$. In particular, we require $V$ to be non-positive within the initial set $X_I$ (\ref{eq:reach_init}) and positive outside the domain \eqref{eq:reach_dom_border} (to ensure we do not leave the domain, where \eqref{eq:reach_deriv} may not hold), while $V$ should be no more negative than a pre-specified level $-\delta<0$ in the rest of the domain $X$  (\ref{eq:reach_else}), and the sublevel set $V$ less than $-\delta$ should be contained within the goal set $X_G$ \eqref{eq:reach_goal}. 
Conditions (\ref{eq:reach_goal})-(\ref{eq:reach_else}) provide a bound on the value of our function which we must reach within the time horizon.

% If $T$ is allowed to tend to infinity (i.e. an infinite time horizon), the difference condition in \eqref{eq:reach_deriv} is reduced to a negativity requirement, as is standard in the literature~\cite{DBLP:conf/hybrid/EdwardsPA24}. 
% Due to our finite time horizon, we require 

The condition in \eqref{eq:reach_deriv} is a decrease condition (its right-hand side is negative due to the choice of $\delta$), that implies $V$ is decreasing along system trajectories till the first time the goal set is reached (by the definition of the time instance $k_G$). 
If $T$ is allowed to tend to infinity (i.e. an infinite time horizon), the difference condition in \eqref{eq:reach_deriv} is reduced to a negativity requirement, as is standard in the literature~\cite{DBLP:conf/hybrid/EdwardsPA24}. 
To gain some intuition on \eqref{eq:reach_deriv}, see that if $k_G = T$, its recursive application leads to 
\begin{align}
V(x(T)) < V(x(0)) - T \frac{1}{T} \Big( \sup_{x \in X_I} V(x) + \delta \Big) \leq -\delta, \label{eq:proof_decr}
\end{align}
where the inequality holds since $V(x(0)) \leq \sup_{x \in X_I}V(x)$. Therefore, if the system starts within $X_I$, then it reaches the goal set (see \eqref{eq:reach_goal}) in at most $T$ steps.

\begin{figure}
	\centering
	\includegraphics[width=.7\linewidth]{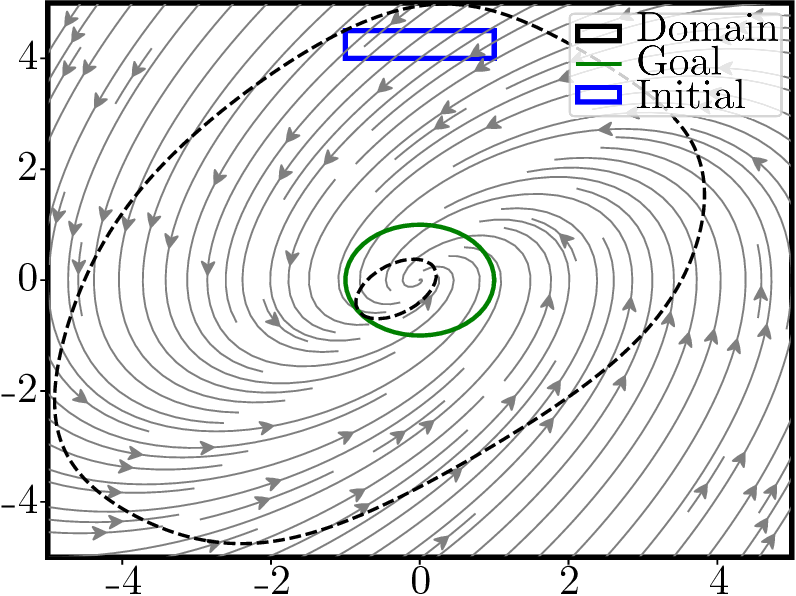}
	\caption{Pictorial illustration of the level sets associated with the reach certificate for the system in \eqref{eq:spiral_dyn}.}
	\label{fig:spiral_reach_plane}
\end{figure}

A graphical representation of these conditions is provided in Figure~\ref{fig:spiral_reach_plane}.
The inner sublevel set (with dashed line) is the set obtained when the certificate value is less than $-\delta$, whilst the outer one is the set obtained when the certificate is less than $0$. 
The decrease condition then means that we never leave the larger sublevel set and must instead converge to the smaller sublevel set.

Now introduce  $\psi^s_{\mathrm{reach}}$ to encode conditions (\ref{eq:reach_init})-(\ref{eq:reach_dom_border}), while
$\psi^\Delta_{\mathrm{reach}}(\xi)$ captures \eqref{eq:reach_deriv}. Notice that the latter depends on $\xi$ as it is enforced on consecutive states $x(k)$ and $x(k+1)$ along a trajectory.

With this in place, we can now define our first certificate.

\begin{prop}[Reachability Certificate]
\label{cert:reach}
    A function $V \colon \mathbb{R}^n \rightarrow \mathbb{R}$ is a reachability certificate if
	 \begin{equation}
	     V \models \psi^s_{\mathrm{reach}} \wedge (\forall \xi \in \Xi) V\models\psi^\Delta_{\mathrm{reach}}(\xi).
	 \end{equation}
\end{prop}
The proof is based on \eqref{eq:proof_decr}; provided formally in Appendix~\ref{app:proofs}.
In words, Proposition~\ref{cert:reach} implies that a function $V$ is a reachability certificate if it satisfies (\ref{eq:reach_init})-(\ref{eq:reach_dom_border}), and (\ref{eq:reach_deriv}) for all trajectories generated by our dynamics.

We now consider a safety property, which is in some sense dual to reachability.

\begin{property}[Safety]\label{prop:safe}
   Consider \eqref{eq:Dyn}, and let $X_I, X_U \subset X$ with $X_I \cap X_U = \emptyset$ denote an initial and an unsafe set, respectively. If for all $\xi \in \Xi$,
    $$
        \phi_\mathrm{safe}(\xi) \defeq \forall k \in \{0,\dots,T\}, x(k) \notin X_U,
    $$
    holds, then we say that $\phi_\mathrm{safe}$ encodes a safety property. $\Xi$ denotes the set of trajectories consistent with \eqref{eq:Dyn} and with initial state contained within $X_I$.
\end{property}
By the definition of $\phi_\mathrm{safe}$, it follows that verifying that a system exhibits the safety property is equivalent to checking that all trajectories emanating from the initial set avoid the unsafe set for all time instances, until horizon $T$. 
The safety property may be constructed for unbounded $X$.

We now define relevant sufficient conditions for a certificate to verify this property, namely, 
\begin{align}
	\label{eq:barr_states1}
    &V(x) \leq 0 , \forall x \in X_I,\\
	\label{eq:barr_states2}
    &V(x) > 0, \forall x \in X_U,\\
        &V(x(k+1))-V(x(k)) \label{eq:barr_deriv} \\ &<\frac{1}{T} \Big (\inf_{x \in X_U}V(x)-\sup_{x \in X_I}V(x) \Big),~ k=0,\dots,T-1. \nonumber
\end{align}
Notice that even if $\inf_{x \in X_U}V(x)-\sup_{x \in X_I}V(x) > 0$, i.e., in the case where the last condition encodes an increase of $V$ along the system trajectories, the system still avoids entering the unsafe set. 
In particular,
\begin{align}
V(x(T)) &< V(x(0)) + \Big (\inf_{x \in X_U}V(x)-\sup_{x \in X_I}V(x) \Big) \nonumber \\
&\leq \inf_{x \in X_U}V(x),
\end{align}
where the inequality holds since $V(x(0)) \leq \sup_{x \in X_I} V(x)$.
Therefore, by \eqref{eq:barr_states2}, the resulting inequality implies that even if the system starts at the least negative state within $X_I$, it will still remain safe. 
Since we consider finite horizon properties, this increase allows us to be less conservative compared with a simple negativity condition, which would be recovered if we allow $T$ to tend to infinity.
% Note that if $T$ tends to infinity (i.e. an infinite time horizon), we recover the standard negativity requirement~\cite{DBLP:conf/hybrid/EdwardsPA24}.}
% \textcolor{blue}{This is in contrast with some of the literature which employs a simple negativity condition~\cite{DBLP:conf/hybrid/EdwardsPA24}, we note that if $T$ tends to infinity (i.e. an infinite time horizon), we recover this negativity requirement, but for a finite $T$ we may be less conservative by allowing an increase.}
% In the case that $T$ tends to infinity (i.e. an infinite time horizon), the difference condition in \eqref{eq:reach_deriv} is reduced to a negativity requirement, as is standard in the literature~\cite{DBLP:conf/hybrid/EdwardsPA24}. 

We denote by $\psi_\text{safe}^s$ the conjunction of \eqref{eq:barr_states1} and \eqref{eq:barr_states2}, and by $\psi^\Delta_\text{safe}(\xi)$ the property in \eqref{eq:barr_deriv}. We then have the following safety/barrier certificate.

\begin{prop}[Safety/Barrier Certificate]
\label{cert:barr}
    A function $V \colon \mathbb{R}^n \rightarrow \mathbb{R}$ is a safety/barrier certificate if
	 \begin{equation}
	     V \models \psi^s_\mathrm{safe} \wedge (\forall \xi \in \Xi) V\models\psi^\Delta_\mathrm{safe}(\xi).
	 \end{equation}
\end{prop}
The proof can be found in Appendix \ref{app:proofs}.
Combining reachability and safety leads to richer properties. One of these is defined next. 

\begin{property}[Reach-While-Avoid (RWA)]\label{prop:rwa}
     Consi-\\der \eqref{eq:Dyn}, and let $X_I, X_U, X_G \subset X$ with $(X_I \cup X_G) \cap X_U = \emptyset$ denote an initial set, an unsafe set, and a goal set, respectively.
     Assume further that $X_G$ is compact and denote by $\partial X_G$ its boundary. 
     If for all $\xi \in \Xi$,
   \begin{equation*}
    \begin{aligned}
                \phi_\mathrm{RWA}(\xi) \defeq \forall k \in \{0,\dots,T\}, x(k) \notin X_U \cup X^c  \\ \wedge~\exists k \in \{0, \dots, T\}, x(k) \in X_G,
    \end{aligned}
    \end{equation*}
    holds, then we say that $\phi_\mathrm{RWA}$ encodes a RWA property.
    $\Xi$ denotes the set of trajectories consistent with \eqref{eq:Dyn} and with initial state contained within $X_I$.
\end{property}
By the definition of $\phi_\mathrm{RWA}$, it follows that verifying that a system exhibits the RWA property is equivalent to verifying that all trajectories emanating from the initial set $X_I$ avoid entering the unsafe set $X_U$ (and the set complement of the domain $X$), and also eventually enter the goal set $X_G$.

The RWA property is derived from the reachability and safety properties, thus the conditions $\psi^s_\mathrm{RWA}$, are the conjunction of $\psi^s_\mathrm{reach}$ and $\psi^s_\mathrm{safe}$, and $\psi^\Delta_\mathrm{RWA}(\xi)$ is given by the conjunction of $\psi^\Delta_\mathrm{reach}(\xi)$ with the following  requirement:  
% Fix $\delta>0$ such that $\delta > -\inf_{x \in X_I} V(x)$.  
% The conditions necessary to verify this property are as follows: 
    \begin{align}
%  \label{eq:RWA_init}
% 		&V(x) \leq 0, \; \forall x \in X_I,\\
%   \label{eq:RWA_safe}
% 		&V(x) > 0, \; \forall x \in X_U,\\
%   \label{eq:RWA_goal}
%   &V(x) \geq -\delta, \; \forall x \in \partial X_G, \\
% 		&V(x) > -\delta, \; \forall x \in X \setminus X_G, 
%   \label{eq:RWA_else}\\
		% &V(x(k+1)) - V(x(k)) \label{eq:RWA_deriv} \\
  %       &<- \frac{1}{T}\left(\sup_{x \in X_I} V(x) +\delta \right),~ k=0,\dots,k_G-1,\nonumber\\
		&V(x(k+1)) - V(x(k)) \label{eq:RWA_deriv2} \\
        &< \frac{1}{T}\left(\inf_{x \in X_U} V(x) +\delta \right),~ k=k_G,\dots,T-1,\nonumber
    \end{align}
% where recall that $k_G$ denotes the first time the system trajectory will ``hit'' the $(-\delta)$-level set of $V$, which is associated with the goal set. 
% We use $\psi^s_\mathrm{RWA}$ to denote the conjunction of (\ref{eq:RWA_init})-(~\ref{eq:RWA_else}), and $\psi^\Delta_\mathrm{RWA}(\xi)$ for \eqref{eq:RWA_deriv} and \eqref{eq:RWA_deriv2}. 

% These conditions ensure that our initial and unsafe sets (including outside the domain) are separated by a zero-level set of the function $V$, and that there is a minimum inside the goal set.
% The difference conditions then ensure that we decrease from the initial set (and hence reach the goal set), and afterward do not increase so much that we enter the unsafe set.

\begin{figure}[b]
    \centering
    \begin{subfigure}{0.3\linewidth}
        \includegraphics[width=.75\linewidth]{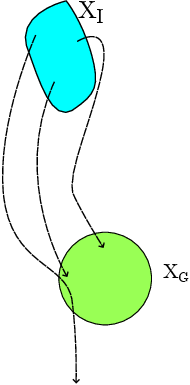}
        \caption{Reachability}
    \end{subfigure}\hfill
        \begin{subfigure}{0.3\linewidth}
        \includegraphics[width=.8\linewidth]{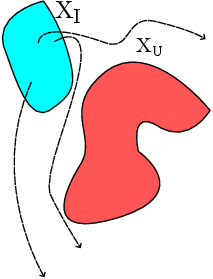}
        \caption{Safety}
    \end{subfigure}\hfill   
    \begin{subfigure}{0.3\linewidth}
        \includegraphics[width=.8\linewidth]{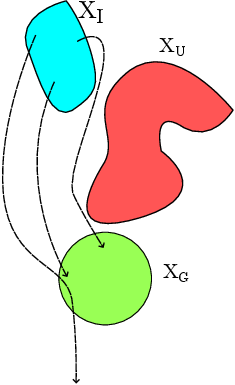}
        \caption{RWA}
    \end{subfigure}
    \caption{Pictorial illustration of (a) reachability, (b) safety, and (c) RWA properties, respectively. Black lines illustrate sample trajectories that satisfy the associated properties.}
    \label{fig:props}
\end{figure}

\begin{prop}[RWA Certificate]
\label{cert:RWA}
    A function $V \colon \\\mathbb{R}^n \rightarrow \mathbb{R}$ is a RWA certificate if
	 \begin{equation}
	     V \models \psi^s_\mathrm{RWA} \wedge (\forall \xi \in \Xi) V\models\psi^\Delta_\mathrm{RWA}(\xi).
	 \end{equation}
\end{prop}
The proof can be found in Appendix \ref{app:proofs}.
We provide a graphical representation of the properties in Figure \ref{fig:props}.

To synthesize one of these deterministic certificates, we require complete knowledge of the behavior $f$ of the dynamical system, to allow us to reason about the space of trajectories $\Xi$.  
This may be impractical, and we therefore consider learning a certificate in a data-driven manner.

\section{Data-Driven Certificates}
\label{sec:learn_certs}

% In some cases, one may have access to the underlying dynamics $f$, in which case it is possible to directly find a certificate that satisfies the relevant criteria $\psi^s$, $\psi^{\Delta}$.
% However, i
% In many real-world scenarios, access to the true dynamics requires a complete model of the physics of the system, and may not always be possible.
% Instead, we take a data-driven approach to synthesize a certificate based only on available trajectories/signatures of that system.

We denote by $(X_I,\mathcal{F},\mathbb{P})$ a probability space, where $\mathcal{F}$ is a $\sigma$-algebra and $\mathbb{P}\colon \mathcal{F}\rightarrow[0,1]$ is a probability measure on the set of initial states $X_I$.
Then, the initial state of the system is randomly distributed according to $\mathbb{P}$.

To obtain our sample set, we consider $N$ initial conditions, sampled from $\mathbb{P}$, namely 
$
    \{x^i(0)\}_{i=1}^N \sim \mathbb{P}^N,$
where we assume that all samples are independent and identically distributed (i.i.d.).
Initializing the dynamics from each of these initial states, we unravel a set of trajectories $\{\xi^i\}_{i=1}^N$.  %$\{ \{x^i(k)\}_{k=0}^T\}_{i=1}^N$. 
Since there is no stochasticity in the dynamics, we can equivalently say that trajectories (generated from the random initial conditions) are distributed according to the same probabilistic law; hence, with a slight abuse of notation, we write $\xi\sim \mathbb{P}$.
In the case of a stochastic dynamical system, the vector field would depend on some additional disturbance vector; our subsequent analysis will remain valid with $\mathbb{P}$ being replaced by the probability distribution that captures both the randomness of the initial state and the distribution of the disturbance. 
 We impose the following mild assumption.
\begin{assum}[Non-concentrated Mass]\label{ass:non-conc_mass}
	Assume that $\mathbb{P}\{\xi \}=0$, for any $\xi \in \Xi$.
\end{assum}

\subsection{Problem Statement}
\label{sec:learn_certs:data}

Since we are now dealing with a sample-based problem, we will be constructing probabilistic certificates and hence probabilistic guarantees on the satisfaction of a given property. We will present our results for a generic property $\phi \in \{\phi_{\mathrm{reach}}, \phi_{\mathrm{safe}}, \phi_{\mathrm{RWA}}\}$ and associated certificate conditions $\psi^s, \psi^{\Delta}$. 

Denote by $V_N$ a certificate of property $\phi$, we introduce the subscript $N$ to emphasize that this certificate is constructed on the basis of sampled trajectories $\{\xi^i\}_{i=1}^N$.

\begin{prob}[Probabilistic Property Guarantee]\label{prob:guarantees}
   Consider $N$ sampled trajectories, and fix a confidence level $\beta \in (0,1)$. We seek a property violation level, or ``risk'', $\epsilon \in (0,1)$ such that
       % \begin{equation}
       % \mathbb{P}^N \big\{ \{\xi^i\}_{i=1}^N \in \Xi^N:
       % \mathbb{P}\{\xi \in \Xi \colon\neg\phi(\xi)\} \leq \epsilon \big \} \geq 1-\beta. \label{eq:prop_prob}%V_N \not\models \psi^s \wedge \psi^\Delta(\xi)\} \leq \epsilon \big \} \geq 1-\beta. \label{eq:prop_prob}
       % \end{equation}
    \begin{align}
       \mathbb{P}^N &\big\{ \{\xi^i\}_{i=1}^N \in \Xi^N:~ \nonumber \\
       &\mathbb{P}\{\xi \in \Xi \colon\neg\phi(\xi)\} \leq \epsilon \big \} \geq 1-\beta. \label{eq:prop_prob}%V_N \not\models \psi^s \wedge \psi^\Delta(\xi)\} \leq \epsilon \big \} \geq 1-\beta. \label{eq:prop_prob}
    \end{align}
\end{prob}

We achieve this by considering a bound on the probability of a new trajectory satisfying our certificate conditions.
Addressing this problem allows us to provide guarantees even if part of the initial set does not satisfy our specification.
Our statement is in the realm of probably approximately correct (PAC) learning: the probability of sampling a new trajectory $\xi \sim \mathbb{P}$ failing to satisfy our certificate condition is itself a random quantity depending on the samples $\{\xi^i\}_{i=1}^N$, and encompasses the generalization properties of a learned certificate $V_N$. It is thus distributed according to the joint probability measure $\mathbb{P}^N$, hence our results hold with some confidence $(1-\beta)$.

Providing a solution to Problem \ref{prob:guarantees} is equivalent to determining an $\epsilon \in (0,1)$, such that with confidence at least $1-\beta$, the probability that $V_N$ does not satisfy the condition $\psi^s \wedge \psi^\Delta(\xi)$ for another sampled trajectory $\xi \in \Xi$ is at most equal to that $\epsilon$. 
As such, with a certain confidence, a certificate $V_N$ \emph{trained} on the basis of $N$ sampled trajectories, will remain a valid certificate with probability at least $1-\epsilon$. 
Therefore, we can argue that  $V_N$ is a \emph{probabilistic} certificate, and hence the property holds (at least) with the same probability.

\subsection{Probabilistic Guarantees}

We now provide a solution to Problem \ref{prob:guarantees}. 
To this end, we refer to a mapping $\mathcal{A}$ such that $V_N = \mathcal{A}(\{\xi^i\}_{i=1}^{N})$ as an algorithm that, based on $N$ samples, returns a certificate $V_N$. Our main result will apply to a generic algorithm that exhibits certain properties outlined as assumptions below. 
In Section \ref{sec:training} we provide a specific synthesis procedure through which $\mathcal{A}$ (and hence the certificate $V_N$) can be constructed, and show that this algorithm satisfies the considered properties.

The following definition constitutes the backbone of our analysis. 
% The notion introduced below appears with different terms in the literature; we adopt the terminology introduced in \cite{DBLP:journals/jmlr/CampiG23,DBLP:journals/tac/MargellosPL15} (adapted to our purposes) to align with the statistical learning literature. 

\begin{defn}[Compression Set]\label{def:compress}
Fix any $\{\xi^i\}_{i=1}^{N}$, and let $\mathcal{C}_N \subseteq \{\xi^i\}_{i=1}^{N}$ be a subset of the samples with cardinality $C_N = |\mathcal{C}_N| \leq N$.
Define $V_{C_N} = \mathcal{A}(\mathcal{C}_N)$.
We say that $\mathcal{C}_N$ is a compression of $\{\xi^i\}_{i=1}^{N}$ for algorithm $\mathcal{A}$, if
    \begin{equation}
    \begin{aligned}
		V_{C_N} = \mathcal{A}(\mathcal{C}_N) = \mathcal{A}(\{\xi^i\}_{i=1}^{N}) = V_N.
    \end{aligned}
    \end{equation}
\end{defn}
Notice the slight abuse of notation, as the argument of $\mathcal{A}$ might be a set of different cardinality; in the following, its domain will always be clear from the context. 

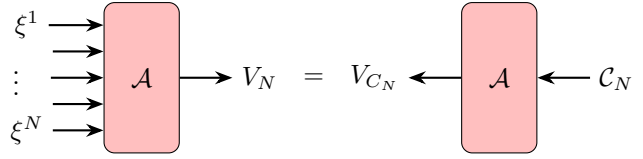
\begin{figure}
    \centering
    \begin{tikzpicture}[node distance=0.35cm ]
        \node (start) {$\xi^1$};
        \node (blank1) [below of=start] {\phantom{$\xi^N$}};
        \node (next) [below of= blank1] {$\vdots$\phantom{$\xi^i$}};
        \node (blank2) [below  of= next] {\phantom{$\xi^N$}};
        \node (final) [below of =blank2] {$\xi^N$};
        
        \node (alg1) [startstop, right=0.7cm of next] {$\mathcal{A}$};
        
        \node (out1) [right=0.7cm of alg1] {$V_N$};
        
        \node (equals) [right = 0.1cm of out1] { $=$};

        \node (out2) [right = 0.1cm of equals] {$V_{C_N}$};

        \node (alg2) [startstop, right=0.7cm of out2] {$\mathcal{A}$};
        
        \node (next2) [right=0.7cm of alg2] {$\mathcal{C}_N$\phantom{$\xi^i$}};
        \node (blank3) [above of=next2] {\phantom{$\xi^N$}};

        \node (blank4) [below of=next2] {\phantom{$\xi^N$}};

        \draw [arrow] (alg1) -- (out1);
        \draw [arrow] (alg2) -- (out2);
        
        \draw [arrow] (start.east) -- (alg1.west|-start.east);
        \draw [arrow] (blank1.east) -- (alg1.west|-blank1.east);
        \draw [arrow] (next.east) -- (alg1.west|-next.east);
        \draw [arrow] (blank2.east) -- (alg1.west|-blank2.east);
        \draw [arrow] (final.east) -- (alg1.west|-final.east);
        \draw [arrow] (next2.west) -- (alg2.east|-next2.west);
        
        \end{tikzpicture}
    \caption{Pictorial illustration of the compression set notion of Definition \ref{def:compress}.}
    \label{fig:Compression}
\end{figure}

Figure \ref{fig:Compression} illustrates Definition \ref{def:compress} pictorially. 
It should be noted that compression set cardinalities may be bounded \emph{a priori}~\cite{DBLP:journals/siamjo/CampiG08}, that is, without knowledge of the sample-set, or obtained \emph{a posteriori}, and hence depending on the given set $\{\xi^i\}_{i=1}^{N}$~\cite{DBLP:journals/mp/CampiG18}. 
We could take a compression set as the entire sample set, resulting in a trivial property violation upper bound of 1. 
    However, it is of benefit to determine a compression set with small (ideally minimal) cardinality, as the smaller $C_N$ is, the smaller risk we can guarantee. 
In this paper we are particularly interested in a posteriori results, since we solve a non-convex problem we cannot in general provide a non-trivial bound to the cardinality of the compression set a priori~\cite{DBLP:journals/tac/CampiGR18}.
Therefore, we introduce the subscript $N$ in our notation for $\mathcal{C}_N$ (set) and $C_N$ (corr. cardinality), respectively.

The properties this algorithm $\mathcal{A}$ must satisfy are as follows (adapted from \cite{DBLP:journals/jmlr/CampiG23}).
% This algorithm $\mathcal{A}$ must satisfy the following properties (adapted from \cite{DBLP:journals/jmlr/CampiG23}; note that in \cite{DBLP:journals/jmlr/CampiG23} a non-concentrated mass property is also imposed, which here appears separately as Assumption \ref{ass:non-conc_mass}). 
% We later demonstrate that our proposed algorithm satisfies these.

\begin{assum}[Properties of $\mathcal{A}$]\label{ass:alg_prop}
Assume that algorithm $\mathcal{A}$ exhibits the following properties:
\begin{enumerate}[wide, labelwidth=!, labelindent=0pt]
	\item \emph{Preference:} For any pair of multisets $\mathcal{C}_1$ and $\mathcal{C}_2$ of elements of $\{\xi^i\}_{i=1}^N$, with $\mathcal{C}_1 \subseteq \mathcal{C}_2$, if  $\mathcal{C}_1$ does not constitute a compression set of $\mathcal{C}_2$ for algorithm $\mathcal{A}$, then $\mathcal{C}_1$ will not constitute a compression set of $\mathcal{C}_2 \cup\{\xi\}$ for any $\xi \in \Xi$.
	\item \emph{Non-associativity:} Let $\{\xi^i\}_{i=1}^{N+\bar{N}}$ for some $\bar{N} \geq 1$. If $\mathcal{C}$ constitutes a compression set of $\{\xi_i\}_{i=1}^{N} \cup \{\xi\}$ for all $\xi \in \{\xi^i\}_{i=N+1}^{N+\bar{N}}$ for algorithm $\mathcal{A}$, then $\mathcal{C}$ constitutes a compression set of $\{\xi_i\}_{i=1}^{N+\bar{N}}$ (up to a measure-zero set).
\end{enumerate}
\end{assum}

If these are satisfied we may use the theorem below to provide probabilistic guarantees on property satisfaction.

\begin{thm}[Probabilistic Guarantees]
\label{thm:Guarantees}
Consider any algorithm $\mathcal{A}$ satisfying Assumption \ref{ass:alg_prop} such that $V_N = \mathcal{A}(\{\xi^i\}_{i=1}^{N})\models \bigwedge_{i=0}^N\psi^\Delta(\xi^i) \wedge \psi^s$, with trajectories $\{\xi^i\}_{i=1}^{N}$ generated in an i.i.d. manner from a distribution satisfying Assumption~\ref{ass:non-conc_mass}. 
Fix $\beta \in (0,1)$, and for $k<N$, let
let $\varepsilon(k,\beta,N)$ be the (unique) solution to the polynomial equation in the interval $[k/N,1]$
    \begin{align}
               \frac{\beta}{2N} \sum_{m=k}^{N-1}&\frac{\binom{m}{k}}{\binom{N}{k}}(1-\varepsilon)^{m-N} \nonumber \\
               &+\frac{\beta}{6N}\sum_{m=N+1}^{4N}\frac{\binom{m}{k}}{\binom{N}{k}}(1-\varepsilon)^{m-N} = 1, \label{eq:eps}
     \end{align}
    while for $k=N$ let $\varepsilon(N,\beta,N) =1$. We then have that
    \begin{align}
	    \label{eq:cert_bound}
        &\mathbb{P}^N\big\{ \{\xi^i\}_{i=1}^N \in \Xi^N:~  \\
        &\mathbb{P}\{\xi \in \Xi\colon \neg \phi(\xi) \} \leq \varepsilon(C_N,\beta,N)\big\} \nonumber \geq 1-\beta.
    \end{align}
\end{thm}
\textbf{Proof}
Fix $\beta \in (0,1)$, and for each $\{\xi^i\}_{i=1}^N$ let $\mathcal{C}_N$ be a compression set for algorithm $\mathcal{A}$. Moreover, note
that letting $V_N = \mathcal{A}(\{\xi^i\}_{i=1}^{N})$ we construct a mapping from samples $\{\xi\}_{i=1}^N$ to a decision, namely, $V_N$, while we impose as an assumption that this mapping satisfies the conditions of Assumption \ref{ass:alg_prop}. 

We first demonstrate that if the certificate conditions are not satisfied on a new sample, then there will be a change in the compression set when the algorithm is fed all samples plus the new violating sample, as follows
\begin{align}
\{\xi \in \Xi &\colon V_N \not\models \psi^s \wedge \psi^\Delta(\xi)) \} \nonumber \\
&\subseteq \{\xi \in \Xi \colon V_N \neq \mathcal{A}(\{\xi\}_{i=1}^N \cup \{\xi\}) \} \nonumber \\
&= \{\xi \in \Xi \colon \mathcal{A}(\mathcal{C}_N) \neq \mathcal{A}(\mathcal{C}_N^{+})\} \nonumber \\
&\subseteq \{\xi \in \Xi \colon \mathcal{C}_N \neq \mathcal{C}_N^{+} \}, \label{eq:proof_thm1}
\end{align}
where $\mathcal{C}_N^{+}$ denotes a compression set for algorithm $\mathcal{A}$ when fed with $\{\xi\}_{i=1}^N \cup \{\xi\}$.
The first inclusion is since for any $\xi \in \Xi$ for which $V_N$ no longer satisfies the certificate condition 
$(\psi^s \wedge \psi^\Delta(\xi))$, we must have that the certificate changes, i.e., $\mathcal{A}(\{\xi^i\}_{i=1}^N \cup \{\xi\})$ (the output of our algorithm when fed with one more sample) is different from $V_N$. 
The opposite statement does not always hold, as having a different certificate does not necessarily mean the old one violates an existing condition for a new $\xi \in \Xi$. 
The equality holds as $V_N = \mathcal{A}(\mathcal{C}_N)$, and $\mathcal{A}(\{\xi\}_{i=1}^N \cup \{\xi\}) = A(\mathcal{C}_N^{+})$, by definition of a compression set.
Finally, the last inclusion stands since any $\xi \in \Xi$ for which $\mathcal{A}(\mathcal{C}_N) \neq \mathcal{A}(\mathcal{C}_N^{+})$, should be such that $\mathcal{C}_N^{+} \neq \mathcal{C}_N$. 
The opposite direction does not always hold, as if 
$\mathcal{C}_N^{+} \supset \mathcal{C}_N$ then we get another compression set of higher cardinality, and hence we may still have $\mathcal{A}(\mathcal{C}_N) = \mathcal{A}(\mathcal{C}_N^{+})$. 

This derivation establishes the fact that the probability of $V_N$ violating the property when it comes to a new $\xi$, is bounded by the probability that the compression set changes, i.e., we have that
\begin{align}
\mathbb{P} \{\xi \in \Xi \colon V_N &\not\models \psi^s \wedge \psi^\Delta(\xi)) \} \nonumber \\
&\leq \mathbb{P}\{\xi \in \Xi \colon \mathcal{C}_N \neq \mathcal{C}_N^{+} \}. \label{eq:proof_thm2}
\end{align}
We can now make use of \cite[Theorem~7]{DBLP:journals/jmlr/CampiG23}, which implies that with confidence at least $1-\beta$, the probability that for a new $\xi \in \Xi$ the compression set changes, is at most $\varepsilon(\mathcal{C}_N,\beta,N)$, i.e., 
\begin{align}
\mathbb{P}\{\xi \in \Xi\colon  \mathcal{C}_N^{+} \neq \mathcal{C}_N\} \leq \varepsilon(C_N,\beta,N), \label{eq:proof_thm}
\end{align}
where the expression of$\varepsilon(k,\beta,N)$ for different values of $k$ is given in \eqref{eq:eps}.
By \eqref{eq:proof_thm2} and \eqref{eq:proof_thm}, 
we have that
\begin{align*}
        &\mathbb{P}^N\big\{ \{\xi^i\}_{i=1}^N \in \Xi^N:~  \\
        &\mathbb{P}\{\xi \in \Xi\colon V_N \not\models \psi^s \wedge \psi^\Delta(\xi))\} \leq \varepsilon(C_N,\beta,N)\big\} \nonumber \geq 1-\beta.
    \end{align*}
By the implication in Definition \ref{prob:property_cert}, \eqref{eq:cert_bound} follows, thus concluding the proof. \qed

The following remarks are in order.
\begin{enumerate}[wide, labelwidth=!, labelindent=0pt]
	\item Notice that Theorem \eqref{thm:Guarantees} involves evaluating $\varepsilon(k,\beta,N)$ at $k=C_N$, i.e., at the cardinality of the compression set. 
    % As such, with confidence at least $1-\beta$ with respect to the choice of the trajectories $\{\xi_i\}_{i=1}^N$, the probability that $V_N$ is not a valid certificate when it comes to a new trajectory $\xi$, is at most $\varepsilon(C_N,\beta,N)$. 
    Due to the dependency of $\varepsilon$ on the samples (via $C_N$), the proposed  probabilistic bound is \emph{a posteriori} as it is adapted to the samples we ``see''. As a result, this is often less conservative compared to \emph{a priori} counterparts.
    \item  
    For cases where algorithm $\mathcal{A}$ takes the form of an optimization program that is convex with respect to the parameter vector, determining non-trivial bounds on the cardinality of compression sets is possible \cite{DBLP:journals/siamjo/CampiG08,DBLP:journals/tac/MargellosPL15}, as this is related to the notion of support constraints in convex analysis.
    However, determining compression sets of low cardinality (necessary for small risk bounds) becomes a non-trivial task if $\mathcal{A}$ involves a non-convex optimization program and/or is iterative (as Algorithm \ref{algo:sub}). 
    This is since in a non-convex setting, samples that give rise to inactive constraints may still belong to a compression set, as they may affect the optimal parameter implicitly. 
    % \textcolor{red}{A direct way of determining the minimal compression set is to resolve the problem with every subset of the samples \cite{DBLP:journals/mp/CampiG18,DBLP:journals/tac/CampiGR18}. 
    % Computationally, however, this would be an intense procedure, often prohibitive due to its combinatorial nature \cite{DBLP:journals/tac/FeleM21}. }
\item An alternative procedure is to use sampled trajectories and to check directly whether a property is satisfied for them (by checking the property definition, rather than using the associated certificate's conditions). 
	This is a valid alternative but has the drawback of not providing a certificate $V_N$, but simply provides an answer as far as the property satisfaction is concerned.
    This direction is pursued in \cite{DBLP:journals/sttt/BadingsCJJKT22}; we review this result and compare with our approach in Section \ref{sec:related:direct_prop}. Note that
	having a certificate is interesting per se, and opens the road for control synthesis, which we aim to pursue in future work. 
\end{enumerate}
\section{Certificate Synthesis}
\label{sec:training}

In this section, we propose mechanisms to synthesize a certificate from sampled trajectories, thus offering a constructive approach for algorithm $\mathcal{A}$ in Theorem \ref{thm:Guarantees}.

% \subsection{Neural Networks}
% In order to learn a certificate from samples, we consider a neural network, a well-studied class of function approximators that generalize well to a given task.

% \begin{defn}[Neural Network]
%     We denote a neural network by an input layer $z_0 \in \mathbb{R}^n$ (same dimension with the system state vector), a number of hidden layers $z_1 \in \mathbb{R}^{h_1}, \dots, z_k \in \mathbb{R}^{h_k}$, and an output layer $z_{k+1} \in \mathbb{R}$.
%     Each layer, except the input, has an associated set of weights and biases $W_i \in \mathbb{R}^{h_{i-1}\times h_i}, b_i \in \mathbb{R}^{h_i}$, as well as an activation function $\sigma_i \colon \mathbb{R} \rightarrow \mathbb{R}$.

%     The layers are related by the following equations,
%     \begin{align}
%         z_{i} &= \sigma_i(W_i z_{i-1}+b_i),~ i =1,\dots,k,\\
%         z_{k+1} &= W_{k+1} z_k+b_{k+1},
%     \end{align}
%     where the activation function is applied element-wise to its argument. 
% \end{defn}

% Such a neural network acts as a ``template'' for our certificate $V_N$. 
We treat a certificate as an appropriately parameterized ``template'' (e.g., neural network), and denote the parameter vector $\theta$. 
We then have that our certificate $V_N$ depends on $\theta$, which is a vector we seek to identify to instantiate our certificate. 
For the results of this section, we simply write $V_\theta$ and drop the dependency on $N$ to ease notation. 

\newcounter{figure_store}
\setcounter{figure_store}{\arabic{figure}}
\setcounter{figure}{0}

\makeatletter
\renewcommand{\fnum@figure}{\textbf{Algorithm \thefigure}}
\makeatother
\begin{figure*}
\caption{Certificate Synthesis and Compression Set Computation}
\vspace{0.2cm}
\label{algo:sub}
\hrule \vspace{0.05cm} \hrule 
    \begin{algorithmic}[1]
    \Function{}{}$\mathcal{A}(\theta,\mathcal{D})$
    \State Set $k \gets 0$ \Comment{Initialize iteration index}
            \State Set $\mathcal{C}\leftarrow \emptyset$ 
            \Comment{Initialize compression set}
            \State Fix $L_1 < L_0$ with $|L_1-L_0|>\eta$ \Comment{$\eta$ is any fixed tolerance}
            \While{$l^s(\theta)>0$} \Comment{While sample-independent state loss is non-zero}
                \State  $g \gets \nabla_\theta l^s(\theta)$  \Comment{Gradient of loss function} \label{line:state_grad}
                \State $\theta \leftarrow \theta-\alpha g$ \Comment{Step in the direction of sample-independent gradient} \label{line:state_step}
            \EndWhile
            \Statex \vspace{-0.35cm}\hrulefill
            \Repeat
            \State $k \gets k+1$ \Comment{Update iteration index} \label{line:update_k}

             % \While{$|L_k-L_{k-1}|>\eta $} \Comment{Iterate until tolerance is met} \label{line:while}
            \State $\mathcal{M} \gets \{\tilde{\xi} \in \mathcal{D} \colon L(\theta,\tilde{\xi}) \geq \max_{\tilde{\xi} \in \mathcal{C}} L(\theta,\tilde{\xi}) \}$ \Comment{Find samples with loss greater than compression set loss} \label{line:max_D}
            \State  $\overline{g}_\mathcal{M} \gets \{\nabla_\theta L(\theta,\tilde{\xi})\}_{\tilde{\xi} \in \mathcal{M}}$  \Comment{ Subgradients of loss function for $\tilde{\xi} \in  \mathcal{M}$} \label{line:subgrad_D}
            \State $\overline{\xi}_{\mathcal{C}} \in \argmax_{\tilde{\xi} \in \mathcal{C}} L(\theta,\tilde{\xi})$ \Comment{Find a sample with maximum loss from $\mathcal{C}$} \label{line:max_C}
            \State $\overline{g}_{\mathcal{C}}  \gets \nabla_\theta L(\theta,\overline{\xi}_{\mathcal{C}})$ \Comment{Approximate subgradient of loss function for $\tilde{\xi} = \overline{\xi}_{\mathcal{C}}$} \label{line:subgrad_C}
            \Statex \vspace{-0.25cm}\hrulefill
             \If{$\exists \overline{g} \in \overline{g}_\mathcal{M} \colon \langle \overline{g},\overline{g}_{\mathcal{C}} \rangle \leq 0 \wedge \overline{g} \neq 0$ \label{line:inner}} \Comment{If there is a misaligned subgradient (take the maximum if multiple)}
                \State $\theta \leftarrow \theta-\alpha\overline{g}$ \Comment{Step in the direction of misaligned subgradient} \label{line:true_step}
                \State $\mathcal{C} \leftarrow \mathcal{C} \cup \{\overline{\xi}\}$ \Comment{Update compression set with sample corresponding to $\overline{g}$} \label{line:update_C}
            \Else
                \State $\theta \leftarrow \theta-\alpha\overline{g}_{\mathcal{C}}$ \Comment{Step in the direction of approximate subgradient} \label{line:approx_step}
            \EndIf \label{line:endif}
	    % \While{$|L_k-L_{k-1}|>\eta $} \Comment{Iterate until tolerance is met} \label{line:while}

     %        \State $\overline{\xi}_\mathcal{D} \in \argmax_{\xi \in \mathcal{D}} L(\theta,\xi)$\Comment{Find a sample with maximum loss from $\mathcal{D}$} \label{line:max_D}
     %        \State  $\overline{g}_\mathcal{D} \gets \nabla_\theta L(\theta,\overline{\xi}_\mathcal{D})$  \Comment{ Subgradient of loss function for $\xi = \overline{\xi}_D$} \label{line:subgrad_D}
     %        \State $\overline{\xi}_{\mathcal{C}} \in \argmax_{\xi \in \mathcal{C}} L(\theta,\xi)$ \Comment{Find a sample with maximum loss from $\mathcal{C}$} \label{line:max_C}
     %        \State $\overline{g}_{\mathcal{C}}  \gets \nabla_\theta L(\theta,\overline{\xi}_{\mathcal{C}})$ \Comment{Approximate subgradient of loss function for $\xi = \overline{\xi}_{\mathcal{C}}$} \label{line:subgrad_C}
     %        \Statex \vspace{-0.25cm}\hrulefill
     %         \If{$\langle \overline{g}_D,\overline{g}_{\mathcal{C}} \rangle \leq 0$} \label{line:inner}
     %         \If{$\overline{g}_D \neq 0$}
     %            \State $\theta \leftarrow \theta-\alpha\overline{g}_\mathcal{D}$ \Comment{Step in the direction of exact subgradient} \label{line:true_step}
     %            \State $\mathcal{C} \leftarrow \mathcal{C} \cup \{\overline{\xi}_D\}$ \Comment{Update compression set with $\xi = \overline{\xi}_D$} \label{line:update_C}
     %            \EndIf
     %        \Else
     %            \State $\theta \leftarrow \theta-\alpha\overline{g}_{\mathcal{C}}$ \Comment{Step in the direction of approximate subgradient} \label{line:approx_step}
     %        \EndIf \label{line:endif}
            \Statex \vspace{-0.35cm}\hrulefill
	    \State $L_k \gets \min\left\{ L_{k-1}, \max_{\xi \in \mathcal{C}} L(\theta,\xi)\right\}$ \Comment{Update ``running'' loss value} \label{line:update_L}
            \Until{$|L_k-L_{k-1}|\leq\eta $} \Comment{Iterate until tolerance is met} \label{line:while}
            % \EndWhile
            \State \Return $\theta, \mathcal{C}_N = \mathcal{C} \cup \argmax_{\xi\in\mathcal{D}}L(\theta,\xi)$
            \EndFunction
    \end{algorithmic}
               \vspace{0.1cm} \hrule \vspace{0.05cm} \hrule
\end{figure*}
\makeatletter
\renewcommand{\fnum@figure}{Fig. \thefigure}
\makeatother

\setcounter{figure}{\arabic{figure_store}}
\setcounter{algorithm}{1}

\subsection{Certificate and Compression Set Computation}

We provide an algorithm that seeks to determine an optimal certificate parameterization $\theta^\star$, resulting in a certificate $V_{\theta^\star}$. 
To this end, for a $\xi \in \Xi$ and parameter vector $\theta$, let \begin{equation}
    L(\theta,\xi)=l^\Delta(\theta,\xi)+ l^s(\theta),\label{eq:opt_prob}
\end{equation} represent an associated loss function consisting of a sample-dependent loss $l^\Delta$, and a sample-independent loss $l^s$. 
Without loss of generality, we assume that we can drive the sample-independent loss to be zero (see further discussions later). 
We impose the next mild assumption, needed to prove termination of our algorithm.
\begin{assum}[Minimizers' Existence] \label{ass:exist}
For any $\{\xi\}_{i=1}^N$, and any non-empty $\mathcal{D} \subseteq \{\xi\}_{i=1}^N$, the set of minimizers of $\max_{\xi \in \mathcal{D}} L(\theta,\xi),$ is non-empty.
\end{assum}

We aim at approximating a minimizer $\theta^\star$ of the quantity $\max_{\xi \in \mathcal{D}} L(\theta,\xi)$ when $\mathcal{D}=\{\xi\}_{i=1}^N$, which exists due to Assumption \ref{ass:exist}. 
We can then use that minimizer to construct $V_{\theta^\star}$. 
To achieve this, we employ Algorithm~\ref{algo:sub}. 
The motivating idea is to perform a subgradient descent step where one is allowed to follow an incorrect gradient as long as it points in the right direction.
We explain the main steps of Algorithm~\ref{algo:sub} with the aid of 
Figure~\ref{fig:algorithm}, where each sample gives rise to a concave triangular constraint.

Algorithm \ref{algo:sub} takes as input some initial (arbitrary) parameter vector $\theta$ and a set of samples $\mathcal{D} \subseteq \{\xi\}_{i=1}^N$. 
First, in steps~\ref{line:state_grad}--\ref{line:state_step}, we optimize by means of a subgradient descent regime for the sample-independent loss until this loss is non-positive, which serves as a form of warm starting.
Then, we follow the subgradient associated with the worst case sample and add it to the compression set $\mathcal{C}$ (step \ref{line:true_step}--\ref{line:update_C}, point $M_1$ in Figure~\ref{fig:algorithm}).
When iterates get to point like $M_2$, the subgradient step becomes inexact, as for the same parameter there exists a sample resulting in a higher loss (see asterisk). 
Such a sample is in $\mathcal{M}$, step~\ref{line:max_D} of Algorithm \ref{algo:sub}. 
However, the algorithm does not ``jump'' to that point, as the inner-product condition in step~\ref{line:inner} of the algorithm is not yet satisfied. 
Graphically, this is since the $M_2$ and the red asterisk are on a side of the respective constraint with the same slope. 
As such the algorithm performs inexact subgradient descent steps up to point $M_3$; this is the first instance where the condition in step~\ref{line:inner} is satisfied (i.e., there exists another constraint with opposite slope\footnote{This constraint with opposite slope may be any constraint with loss greater than the loss evaluated on the compression set, not just the maximum one.}) and hence the algorithm ``jumps'' to a point with higher loss and subgradient of opposite sign. 
This procedure is then repeated as shown in the figure, with the red line indicating the iterates' path.
The ``jumps'' serve as an exploration step to investigate the non-convex landscape, while their number (plus two for initial and final worst case sample) corresponds to the cardinality of the returned compression set. 
We iterate till the loss value meets a given tolerance $\eta$ (see steps~\ref{line:while} and~\ref{line:update_L}).
It is to be understood that if $\mathcal{C}$ is empty (as per initialization) steps~\ref{line:max_C}-\ref{line:subgrad_C} are not performed.

\begin{figure}[h!]
\centering
\begin{tikzpicture}[>=latex, scale = 1]   
%==================
% left black
\draw[line width=1.25pt] plot[domain=-4.2:-1] (\x,{-1*(\x + 4.2)+3.2});      
\draw[line width=1.25pt] plot[domain=-5.3:-4.2] (\x,{1*(\x + 4.2)+3.2});   
\fill[pattern=vertical lines] (-4.2,3.2) -- (-1,0) -- (-1,-0.3) -- (-4.2,2.9) -- cycle;      
\fill[pattern=vertical lines] (-5.3,2.1) -- (-4.2,3.2) -- (-4.2,2.9) -- (-5.3,1.8) -- cycle;

% middle black
\draw[line width=1.25pt] plot[domain=-3.4:-0.93] (\x,{-1.6*(\x + 3.4)+3.96});      
\draw[line width=1.25pt] plot[domain=-5.3:-3.4] (\x,{1.6*(\x + 3.4)+3.96});   
\fill[pattern=vertical lines] (-3.4,3.96) -- (-0.93,0) -- (-0.93,-0.3) -- (-3.4,3.66) -- cycle;      
\fill[pattern=vertical lines] (-5.3,0.92) -- (-3.4,3.96) -- (-3.4,3.66) -- (-5.3,0.62) -- cycle;

% right black
\draw[line width=1.25pt] plot[domain=-0.5:0.9] (\x,{-0.85*(\x + 0.5)+3});      
\draw[line width=1.25pt] plot[domain=-4.02:-0.5] (\x,{0.85*(\x + 0.5)+3});   
\fill[pattern=vertical lines] (-0.5,3) -- (0.9,1.81) -- (0.9,1.51) -- (-0.5,2.7) -- cycle;      
\fill[pattern=vertical lines] (-4.02,0) -- (-0.5,3) -- (-0.5,2.7) -- (-4.02,-0.3) -- cycle;

% inactive
\draw[line width=1.25pt] plot[domain=0.2:0.9] (\x,{-2.2*(\x - 0.2)+3.7});      
\draw[line width=1.25pt] plot[domain=-1.48:0.2] (\x,{2.2*(\x - 0.2)+3.7});   
\fill[pattern=vertical lines] (0.2,3.7) -- (0.9,2.16) -- (0.9,1.86) -- (0.2,3.4) -- cycle;      
\fill[pattern=vertical lines] (-1.48,0) -- (0.2,3.7) -- (0.2,3.4) -- (-1.48,-0.3) -- cycle;

% 
%==================
%==================
% left panel
\draw[->,thick] (-5.5,0)--(1.5,0) node[below]{};
\draw[->,thick] (-5.3,-0.2)--(-5.3,4.5) node[left]{Loss};
\draw[->,thick, red!80] (-4.1,3.25)-- (-3.25,2.4) node[below]{};
\draw[->,thick, red!80] (-3.25,2.4) -- (-2.4,1.55) node[below]{};
\draw[->,thick, red!80] (-2.4,1.55) -- (-1.55,0.7) node[below]{};
\draw[->,thick, red!80] (-1.55,0.7) --  (-1.55,2.25) node[below]{};
\draw[->,thick, red!80] (-1.55,2.25) -- (-2.2,1.7) node[below]{};
\draw[->,thick, red!80] (-2.2,1.7) -- (-2.2,2.25) node[below]{};
\draw[->,thick, red!80] (-2.2,2.25) -- (-1.95,1.85) node[below]{};
\filldraw[black] (-4.1,3.25) circle (2pt);
\filldraw[black] (-3.25,2.4) circle (2pt);
\filldraw[black] (-2.4,1.55) circle (2pt);
\filldraw[black] (-1.55,0.7) circle (2pt);
\filldraw[black] (-1.55,2.25) circle (2pt);
\filldraw[black] (-2.2,1.7) circle (2pt);
\filldraw[black] (-2.2,2.25) circle (2pt);
\draw[] (0.2,-0.2)node[below, bag]{Parameter Space};
\draw[]  (-4.1,3.3)node[above, bag]{$M_1$};
\draw[]  (-3.25,2.5)node[above, bag]{$M_2$};
\draw[red!80] (-3.25,3.8) node[above, bag]{$\large{\boldsymbol{\ast}}$};
\draw [->,ultra thick] (0,1.2) to [bend right=30] (-1.5,0.8);
\draw[]  (0.25,0.9) node[above, bag]{$M_3$};
%==================
\end{tikzpicture}
\caption{Graphical illustration of Algorithm~\ref{algo:sub}.}
\label{fig:algorithm}
\end{figure}

% \begin{figure}[b]
%     \centering
%     \includegraphics[width=0.75\linewidth]{Figures/Algorithm.eps}
%     \caption{Graphical Representation of Algorithm~\ref{algo:sub}.}
%     \label{fig:algorithm}
% \end{figure}

Overall, Algorithm~\ref{algo:sub} can be viewed as a specific choice for the mapping $\mathcal{A}$ introduced in Section \ref{sec:learn_certs} when fed with $\mathcal{D} = \{\xi_i\}_{i=1}^N$, and some initial choice for $\theta$. 
It follows a subgradient descent scheme with ``jumbs'' that (i) allows minimizing a (possibly) non-convex loss function, and (ii) the mechanism that triggers the ``jumps'' provides the means to compute 
a compression set. Such a mechanism serves as an efficient alternative to existing methodologies, as we construct it iteratively. 
At the same time the constructed compression set is non-trivial as we avoid adding uninformative samples to it, and only add one sample per iteration in the worst case.
However, the added sample has a loss higher than that of the compression samples (see step~\ref{line:max_D}), and is also informative in the sense of having a misaligned subgradient that allows for exploration (see step~\ref{line:inner}). It should be highlighted that the underpinning idea of constructing the compression set by incrementing it by one sample at a time is inspired by the so called pick-to-learn paradigm proposed in~\cite{DBLP:conf/nips/PaccagnanCG23}. That methodology is general and does not involve a gradient-descent scheme equipped with our proposed logic. This design is thus novel and serves as a constructive instance of the general methodology of~\cite{DBLP:conf/nips/PaccagnanCG23}. 

The main features of Algorithm \ref{algo:sub} are summarized in the proposition below, while its proof can be found in Appendix~\ref{app:proofs}.

\begin{prop}[Algorithm \ref{algo:sub} Properties] \label{prop:converge}
Consider Assumption \ref{ass:non-conc_mass}, Assumption  \ref{ass:exist} and Algorithm \ref{algo:sub} with $\mathcal{D} = \{\xi_i\}_{i=1}^N$ and a fixed (sample independent) initialization for the parameter $\theta$. We then have:
\begin{enumerate}[wide, labelwidth=!, labelindent=0pt]
\item Algorithm \ref{algo:sub} terminates, returning a parameter vector $\theta^\star$ and a set $\mathcal{C}_N$.
\item The set $\mathcal{C}_N$ with cardinality $C_N = |\mathcal{C}_N|$ forms a compression set for Algorithm \ref{algo:sub}.
\item Algorithm \ref{algo:sub} satisfies Assumption \ref{ass:alg_prop}.
% \item \textcolor{blue}{Algorithm~\ref{algo:sub} will converge to the (local) optimum of the pointwise maximum of (locally) convex loss functions.}
\end{enumerate}
\end{prop}

% The proof can be found in Appendix \ref{app:proofs}.

Proposition \ref{prop:converge} implies that we can construct a certificate $V_N = V_{\theta^\star}$, while the algorithm that returns this certificate satisfies Assumption \ref{ass:alg_prop} and admits a compression set $\mathcal{C}_N$ with cardinality $C_N$. 
As such, Algorithm \ref{algo:sub} offers a constructive mechanism to synthesize a certificate, and, if the loss is driven to zero (the assumed minimum value), then all certificate conditions are met and hence the probabilistic guarantees obtained refer to guarantees on the probability of satisfaction of the underlying property.
It should also be noted that in in the numerical simulations presented below, we equipped the subgradient descent scheme with a momentum term thus constructing a deterministic version (as the step size is deterministic) of the so called Adam algorithm~\cite{DBLP:journals/corr/KingmaB14} to boost performance. 
 %can be accompanied by the probabilistic guarantees of Theorem \ref{thm:Guarantees}. 
% Moreover, Assumptions \ref{ass:non-conc_mass} \& \ref{ass:exist} under which Algorithm \ref{algo:sub} exhibits these properties are rather mild.

% This algorithm could be thought of as a constructive procedure for the general methodology proposed recently in \cite{DBLP:conf/nips/PaccagnanCG23}.

% \textcolor{blue}{By construction, Algorithm \ref{algo:sub} creates a non-increasing sequence of iterates $\{L_k\}_{k\geq 0}$ that is bounded below by the global minimum of $\min_{\xi \in D} L(\theta,\xi)$ which exists and is finite due to Assumption \ref{ass:exist}.}
% As such, the algorithm moves towards the direction of the optimum, but we have no guarantees that it indeed reaches some (local) optimum.
% We conjecture the approximate subgradient used in our algorithm constitutes a descent direction~\cite{doi:10.1137/1.9781611971309}, and hence if the step size is chosen appropriately the algorithm should converge to a stationary point. 
% Current work focuses on formalizing this claim.
% \textcolor{blue}{Indeed, for a (locally) convex loss function we can guarantee convergence to within a region of the optimum within the convex region.}

% \color{blue}{
% \begin{prop}[Algorithm~\ref{algo:sub} Converges
    
% \end{prop}
% }
% \end{rem}

\subsection{Discarding Mechanism} 

In some cases, the parameter returned by Algorithm \ref{algo:sub} may result in a value of the loss function that is considered as undesirable (and as a result the constructed certificate might be far from meeting the desired conditions). 
To achieve a lower loss, we make use of a sample-and-discarding procedure~\cite{DBLP:journals/jota/CampiG11,DBLP:journals/tac/RomaoPM23}.
To this end, consider Algorithm \ref{algo:main}. At each iteration of this algorithm, the compression set returned by Algorithm \ref{algo:sub} (step~\ref{line:subgrad}) is discarded from $\mathcal{D}$, and added to a record of the set of removed samples $\mathcal{R}$ (steps~\ref{line:discard}--\ref{line:update_outer_C}). 
We repeat the process till the worst case loss $\max_{\xi \in \mathcal{D}} L(\theta, \xi)\geq0$ becomes zero (its minimum value). 
This implies that Algorithm \ref{algo:sub} is invoked each time with fewer samples as its input, while the set $\mathcal{R}$ progressively increases.
The set of samples that are removed across the algorithm's iterations is denoted by $\mathcal{R}_N$, and forms a compression set for Algorithm~\ref{algo:main}. 
However, it has higher cardinality compared to the original compression set, implying that improving the loss comes at the price of an increased risk level $\varepsilon$ as the cardinality of the compression set increases.

\begin{algorithm}[ht]
\caption{Compression Set Update with Discarding}
\vspace{0.2cm}
\label{algo:main}
\hrule \vspace{0.05cm} \hrule \vspace{0.1cm}
\begin{algorithmic}[1]
\State Fix $ \{\xi^i\}_{i=1}^N$
    \State Set $\widetilde{\mathcal{C}}\gets \emptyset$\Comment{Initialize compression set}
    \State Set $\mathcal{D} \gets \{\xi^i\}_{i=1}^N$ \Comment{Initialize ``running'' samples}
    %\State $\rhd$ While loss is positive
    \While{$\max_{\xi \in \mathcal{D}} L(\theta, \xi)>0$}
            \State $\theta, \mathcal{C} \gets$ $\mathcal{A}(\theta,\mathcal{D})$ \Comment{Call Algorithm \ref{algo:sub}} \label{line:subgrad}
            \State  $\mathcal{D} \gets \mathcal{D} \setminus \mathcal{C}$ \Comment{Discard compression set $\mathcal{C}$ from $\mathcal{D}$} \label{line:discard}
            \State $\mathcal{R} \gets \mathcal{R} \cup \mathcal{C}$ \Comment{{Store discarded samples}} \label{line:update_outer_C}
    \EndWhile
        \State \Return $\theta$, $\mathcal{R}_N=\mathcal{R}$
\end{algorithmic}
\vspace{0.1cm}
\hrule \vspace{0.05cm} \hrule 
\end{algorithm}

This algorithm can be thought of as an add-on to Algorithm \ref{algo:sub} and in general to the procedure of~\cite{DBLP:conf/nips/PaccagnanCG23} as it offers the means to trade the size of the compression set to performance. Unlike Algorithm \ref{algo:sub} and \cite[Algorithm 1]{DBLP:conf/nips/PaccagnanCG23} that iteratively increase the samples used for learning, Algorithm \ref{algo:main} gradually decreases the number of samples used as input to $\mathcal{A}$ across iterations.

%We also provide a novel algorithm to learn a compression set at each iteration (Algorithm~\ref{algo:sub}), with favourable properties for compression set computation, which was not explored in~\cite{DBLP:conf/nips/PaccagnanCG23}.}

\begin{prop}[Algorithm \ref{algo:main} Properties] \label{prop:converge_main}
Consider Assumption \ref{ass:non-conc_mass}, Assumption  \ref{ass:exist} and Algorithm \ref{algo:main} with $\mathcal{D} = \{\xi_i\}_{i=1}^N$ and a fixed (sample independent) initialization for the parameter $\theta$. We then have:
\begin{enumerate}[wide, labelwidth=!, labelindent=0pt]
\item Algorithm \ref{algo:main} converges to a minimum loss value of zero, returning a parameter vector $\theta^\star$ and a set $\mathcal{R}_N$.
\item The set $\mathcal{R}_N$ with cardinality $R_N = |\mathcal{R}_N|$ forms a compression set for Algorithm \ref{algo:main}.
\item Algorithm \ref{algo:main} satisfies Assumption \ref{ass:alg_prop}.
% \item \textcolor{blue}{Algorithm~\ref{algo:sub} will converge to the (local) optimum of the pointwise maximum of (locally) convex loss functions.}
\end{enumerate}
\end{prop}
%Hence, we can make use of Algorithm~\ref{algo:main}, to learn a certificate with guarantees from Theorem~\ref{thm:Guarantees}.}

The proof of this proposition can be found in Appendix~\ref{app:proofs}. Since it establishes that Algorithm \ref{algo:main} satisfies Assumption \ref{ass:alg_prop}, we have that Algorithm \ref{algo:main}  enjoys the guarantees of Theorem~\ref{thm:Guarantees} with $\mathcal{R}_N$ in place of $\mathcal{C}_N$.

The cardinality of the compression set does not necessarily increase with the state space dimension, but is rather dependent on the complexity of the problem.  
For example, a problem where some trajectories approach or even enter the unsafe set presents a more challenging synthesis problem than one where trajectories all move in the opposite direction to the unsafe set, thus we expect the former to have a larger compression set even if the problem is smaller in dimension. This claim is supported numerically by the results of Section \ref{sec:exp}.
% Trajectories that enter the unsafe set will eventually be discarded, leading to a larger compression set and a higher risk bound (as expected for a partially unsafe system).

\subsection{Choices of Loss Function}
We now provide some choices of the loss function $L(\theta,\xi)=l^\Delta(V_\theta, \xi) + l^s(V_\theta)$ so that minimizing that function we obtain a parameter vector $\theta^\star$, and hence also a certificate $V_{\theta^\star}$, which satisfies the conditions of the property under consideration, namely, reachability, safety, or RWA.
Note that when calculating subgradients to these functions, which as we will see below are non-convex, we effectively have the so-called Clarke subdifferential~\cite{doi:10.1137/1.9781611971309}.

We provide some expressions for $l^s$ and $l^\Delta$ for the reachability property in Property \ref{prop:reach}. 
For the other properties, the loss functions can be defined in an analogous manner. 
To this end, we define
\begin{align}
    &l^s(V_\theta) \defeq \int_{X \setminus X_G}\max\{0,-\delta-V_\theta(x)\} ~\mathrm{d}x \\ &+\int_{X_I}\max\{0,V_\theta(x)\}   ~\mathrm{d}x+\int_{\mathbb{R}^N \setminus X} \max\{0,-V_\theta(x)\}  ~\mathrm{d}x.\nonumber
\end{align}
Focusing on the first of these integrals, if $V(x) > -\delta$ then $\max\{0, -\delta-V_\theta(x)\}=0$, i.e., no loss is incurred, implying satisfaction of \eqref{eq:reach_goal}, \eqref{eq:reach_else}. 
Under a similar reasoning, the other integrals account for \eqref{eq:reach_init} and \eqref{eq:reach_dom_border}, respectively. 
For a sufficiently expressive function approximator, we can find a certificate $V$ which satisfies the state constraints and hence has a sample-independent loss of zero.

In practice, we replace integrals with a summation over points generated deterministically within the relevant domains. 
These points are generated densely enough across the domain of interest, and hence offer an accurate approximation. 
This generation may happen through gridding the relevant domain, or sampling according to a fixed synthetic distribution; these samples are considered here to be fixed and they are not related with the ones used to provide probabilistic guarantees.
For the last term, we only enforce the positivity condition on the border of the domain $X$.
Thus, we take a deterministically generated discrete set of points on each domain $\mathcal{X}_{\overline{G}}$ for points in the domain but outside the goal region, $\mathcal{X}_I$ from the initial set, and $\mathcal{X}_\partial$ for the border of the domain $X$.
% Since these samples do not require access to the dynamics we consider them separate to the sample-set $\{\xi_i\}_{i=1}^N$, and references to the size of the sample set only refer to the trajectory samples (since these are the ``costly'' samples).
Our loss function takes then the form: 
\begin{align}
    &\hat{l^s}(V_\theta) \defeq  \frac{1}{|\mathcal{X}_{\overline{G}}|}\sum_{x \in \mathcal{X}_{\overline{G}}} \max\{0, -\delta-V_\theta(x)\} \\ &+\frac{1}{|\mathcal{X}_I|}\sum_{x \in \mathcal{X}_I}\max\{0,V_\theta(x)\}+\frac{1}{|\mathcal{X}_\partial|}\sum_{x \in \mathcal{X}_\partial} \max\{0,-V_\theta(x)\}.\nonumber
    \end{align}

We define $l^\Delta$ by
\begin{equation}
\label{eq:c_deriv}
        \begin{aligned}            
        l^\Delta&(V_\theta, \xi) \defeq \max \Big \{ 0, \max_{k=0,\dots,k_G-1} \Big ( V_\theta(x(k+1))-V_\theta(x(k)) \Big )\\&-\frac{1}{T} \Big (\sup_{x\in \mathcal{X}_I} V_\theta(x) + \delta \Big ) \Big \}.
        \end{aligned}
\end{equation}

The value of $l^\Delta$ encodes a loss if the condition in \eqref{eq:reach_deriv} is violated.
If both $l^s$ and $l^\Delta$ evaluate to zero for all $\{\xi\}_{i=1}^N$, then we have that 
\begin{equation}
\begin{aligned}
        l^s(V_\theta) + \max_{i = 1, \dots, N} l^\Delta(V_\theta, \xi^i) = 0,
\end{aligned}
\end{equation}
which by Certificate \ref{cert:reach} implies that the constructed certificate $V_\theta$ is such that
\begin{equation}
\begin{aligned}
        V_\theta \models \psi^s_\text{reach} \wedge (i=1,\dots,N) V_\theta \models\psi^\Delta_\text{reach}(\xi^i).
\end{aligned}
\end{equation}
Analogous conclusions hold for all other certificates.
\section{Comparison with Related Work}
\label{sec:related}
\subsection{Direct Property Evaluation}
\label{sec:related:direct_prop}
As is known in the case of Lyapunov stability theory, the existence of a certificate is useful per se, and allows one to translate a property to a scalar function. 
% As discussed in Corollary~\ref{corr:prop}, a byproduct of this certificate synthesis is that they provide guarantees on the probability of property violation (see \eqref{eq:prop_viol}). 
However, if one is not interested in the construction of a certificate and only in such guarantees, then Theorem 2 in \cite{DBLP:journals/sttt/BadingsCJJKT22} provides an alternative. 
% We adapt this result in the proposition below, using the Langford binomial tail bound~\cite{JMLR:v6:langford05a} to obtain a tighter guarantee than the sampling-and-discarding result~\cite{DBLP:journals/jota/CampiG11} used in \cite{DBLP:journals/sttt/BadingsCJJKT22}. 
\begin{prop}[Theorem $2$ in \cite{DBLP:journals/sttt/BadingsCJJKT22}\footnotemark]
\label{corr:a_post}
Fix $\beta \in (0,1)$, and for $r = 0,\ldots,N-1$,
determine $\varepsilon(r,\beta,N)$ such that 
    \begin{equation}   
    \label{eq:eps_orig}
     \sum_{k=0}^r\binom{N}{k} \varepsilon^k(1-\varepsilon)^{N-k}=\frac{\beta}{N},
    \end{equation}
  while for $r=N$ let $\varepsilon(N,\beta,N) = 1$.   
	Denote by $R_N$ the number of samples in $\{\xi^i\}_{i =1}^{N}$ for which $\phi(\xi^i)$ is violated.
    We then have that
    \begin{align}
	    \mathbb{P}^N &\big \{ \{\xi^i\}_{i=1}^N \in \Xi^N:~\nonumber \\
        &\mathbb{P}\{\xi \in \Xi \colon \neg\phi(\xi) \} \leq \varepsilon(R_N,\beta,N)\big\}
        \geq 1-\beta. \label{eq:prop_direct}
    \end{align}
\end{prop}
%\footnotetext{\color{blue}Although not considered in~\cite{DBLP:journals/sttt/BadingsCJJKT22}, these bounds may be improved by using the Langford binomial tail bound~\cite{JMLR:v6:langford05a}, replacing $\frac{\beta}{N}$ with $\beta$.}
% This involves a direct application of the Langford binomial tail bound~\cite{JMLR:v6:langford05a}. % (offering an improvement on the sampling-and-discarding results in \cite{DBLP:journals/jota/CampiG11} which require $\frac{\beta}{N}$ on the right-hand side of \eqref{eq:prop_direct}). 
This is an \emph{a posteriori} result, as $R_N$ can be determined only once the samples are observed. 
% To this end, the term $\beta /N$ appears in the right-hand side of \eqref{eq:prop_direct} to account for the fact that, depending on the samples, up to $N$ terms could appear in the summation.
In this case, we have a compression set which is the set of all discarded samples, plus an additional one to support the solution after discarding.
Since this additional sample is always present, we incorporate it in the formula in \eqref{eq:eps_orig}. 

We remark that one could obtain  different bounds through alternative statistical techniques, such as Hoeffding's inequality~\cite{Hoeffding01031963} or Chernoff's bound~\cite{10.1214/aoms/1177729330}. 
Since these bounds are of different nature, we do not pursue that avenue further here.

%For property violation Hoeffding bounds expected value (hence violation rate) which is the same as what scenario approach bound does because we have a bernoulli distribution. 
% Hence no reason we couldn't use Hoeffding, maybe we could also apply it to the result from certificate synthesis? Then the below arguments hold.

% However, such techniques have the drawback of only probabilistically bounding a distance from the mean, rather than providing a bound on the probability of drawing a violating sample.}
% \textcolor{red}{[next statement unclear:]}
% Further, one requires bounds on the loss $L(\theta, \xi)$ and hence on the certificate value and Lie derivative, which may not be available.}

% \textcolor{red}{[pls rephrase, very clear:]}

\begin{figure}
    \centering
    \includegraphics[width=.8\linewidth]{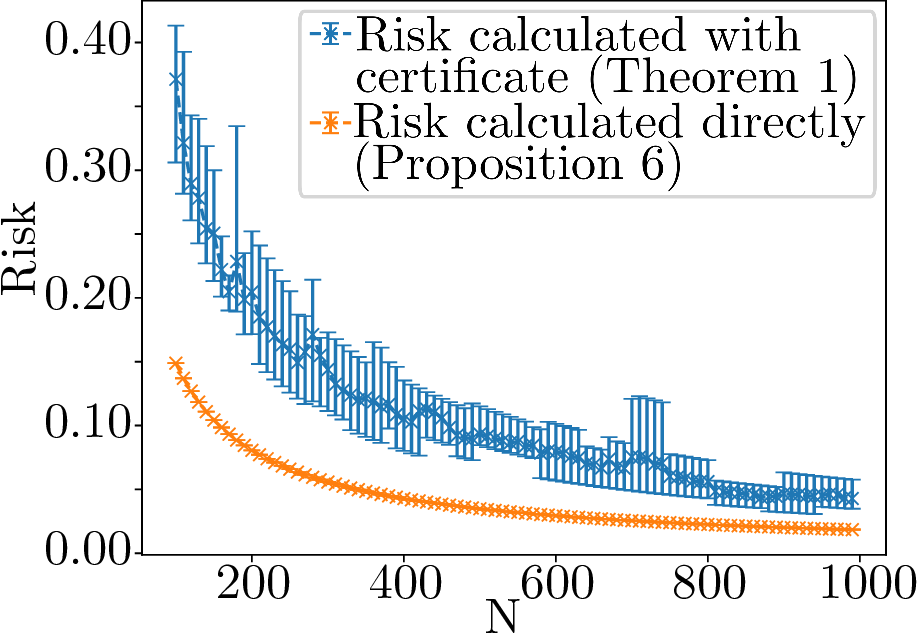}
    \caption{Comparison of the bounds in Theorem~\ref{thm:Guarantees} and Proposition~\ref{corr:a_post} for direct property evaluation. Median values across the 5 runs are shown with a cross, and ranges are indicated by error bars.}
    \label{fig:bounds}
\end{figure}

% Proposition \ref{corr:a_post} offers an alternative to Theorem \ref{thm:Guarantees} to directly bound property violation. 
We compare the risk levels $\varepsilon$ computed by each approach on a benchmark example in \eqref{eq:spiral_dyn} under a \textit{safety} specification; general conclusions are case dependent, as both bounds are \emph{a posteriori}.
For a fixed $\beta$, Figure~\ref{fig:bounds} shows the resulting risk levels for varying $N$ across 5 independently sampled sets of trajectories.
The difference between the orange curve and the blue one can be interpreted as the price related to certificate generation, as per Theorem \ref{thm:Guarantees}. For sufficiently large $N$, this price is marginal.  
As the specification is deterministically safe, no discarding is performed for Proposition~\ref{corr:a_post}, resulting in a smooth curve without variability. 
For non-zero $R_N$ we expect variability as $R_N$ will be randomly distributed. 

%Theorem 5.3
\subsection{Certificate Synthesis as in \cite{DBLP:journals/tac/NejatiLJSZ23}}
The results in \cite{DBLP:journals/tac/NejatiLJSZ23} constitute the closest to our work. 
As no results on reachability and RWA problems were provided in \cite{DBLP:journals/tac/NejatiLJSZ23}, we limit our discussion to the \textit{safety} property. 
As with our work, a sample-based construction is performed, where samples therein are pairs (state, next-state), as opposed to trajectories as in our work. However, the probabilistic bounds established in \cite{DBLP:journals/tac/NejatiLJSZ23} are structurally different and of complementary nature to our work: next, we review the main result in \cite{DBLP:journals/tac/NejatiLJSZ23}, adapted to our notation.

\begin{thm}[Theorem $5.3$ in~\cite{DBLP:journals/tac/NejatiLJSZ23}]
    Consider \eqref{eq:Dyn}, with initial and unsafe sets $X_I,X_U\subset X \subset \mathbb{R}^n,$ respectively.
    Consider also $N$ samples $\{x_i, f(x_i)\}_{i=1}^N$ from $X$, and assume that the loss function in \eqref{eq:opt_prob} is Lipschitz continuous with constant $\mathcal{L}$.
    Consider then the problem
        \begin{align}
            \eta^\star_N &\in \argmin_{d = (\gamma,\lambda,c,\theta),\eta \in \mathbb{R}}\eta \nonumber \\
            \text{st. }&\; V_\theta(x)-\gamma \leq \eta, \; \forall x \in X_I \nonumber \\
            &\; V_\theta(x)-\lambda \geq -\eta, \; \forall x \in X_U \nonumber \\
            &\; \gamma + cT - \lambda - \mu \leq \eta, ~ c \geq 0, \nonumber \\
            &\; V_\theta(f(x_i)) - V_\theta(x_i) -c \leq \eta, \; i=1,\dots,N,  
        \end{align}
        where $\theta$ parameterizes $V_\theta$, and all other decision variables are scalars leading to level sets of $V_\theta$. Let $\kappa(\delta)$ be such that
    \begin{equation}
	    \label{eq:ball}
        \kappa(\delta) \leq \mathbb{P}\{\mathbb{B}_\delta(x)\}, \forall \delta \in \mathbb{R}_{\geq 0}, \forall x \in X,
    \end{equation}
    where $\mathbb{B}_\delta(x) \subset X$ is a ball of radius $\delta$, centered at $x$.
    Fix $\beta \in (0,1)$ and
    determine $\epsilon(|d|,\beta,N)$ from \eqref{eq:eps_orig}, with $r = d$ and by replacing the right hand-side with $\beta$.
    If $\eta^\star_N \leq \mathcal{L} \kappa^{-1}(\epsilon(|d|,\beta,N)$, we have that
    \begin{equation}
	    \mathbb{P}^N\big \{ \{\xi^i\}_{i=1}^N \in \Xi^N:~\phi_{\mathrm{safe}}(\xi), \; \forall \xi \in \Xi  \big\} \geq 1-\beta.
    \end{equation}
\end{thm}

The following remarks are in order.
\begin{enumerate}[wide, labelwidth=!, labelindent=0pt]
	\item The result in \cite{DBLP:journals/tac/NejatiLJSZ23}, capitalizing on the developments of \cite{6832537}, is \emph{a priori}, as opposed to the \emph{a posteriori} assessments of our analysis that are in turn based on \cite{DBLP:journals/jmlr/CampiG23}.
    Moreover, \cite{DBLP:journals/tac/NejatiLJSZ23} offers a
    guarantee that, with a certain confidence, the safety property is \emph{always} satisfied. This is in contrast to Theorem \ref{thm:Guarantees} where we provide such guarantees in probability (up to a quantifiable risk level $\varepsilon$).
However, these ``always'' guarantees come with potential challenges. In particular, the constraint in \eqref{eq:ball} involves the measure of a ``ball'' in the uncertainty space. 
The measure of this ball grows exponentially in the dimension of the uncertainty space (see also Remark 3.9 in 
    \cite{6832537}), while it depends linearly on the dimension of the decision space $|d|$ (see dependence of $\varepsilon$ below \eqref{eq:ball}). This dependence in the results of \cite{DBLP:journals/tac/NejatiLJSZ23} raises computational challenges to obtain useful bounds: we demonstrate this numerically in Section~\ref{sec:exp} employing  one of the examples considered in \cite{DBLP:journals/tac/NejatiLJSZ23}. On the contrary, Theorem \ref{thm:Guarantees} depends only on the cardinality of the compression set. 
	\item The result in \cite{DBLP:journals/tac/NejatiLJSZ23} requires inverting $\kappa(\delta)$, which may not have an analytical form in general. Moreover, it implicitly assumes some knowledge of the distribution to obtain $\kappa$, and of the Lipschitz constants of the system dynamics, which we do not require in our analysis. 
\item The results of \cite{DBLP:journals/tac/NejatiLJSZ23} can be extended to continuous-time dynamical systems, which is also possible for our results but outside the scope of this article: we refer to~\cite{DBLP:journals/corr/abs-2503-13392} for extensions to such cases.
%however, due to practical considerations, we then require knowledge of the Lipschitz constants not required in the discrete time setting. 
%This discussion is not pursued further here. 
\end{enumerate}

\section{Numerical Results}
\label{sec:exp}

\blfootnote{The codebase is available at \url{https://github.com/lukearcus/fossil_scenario}}

% \footnotetext{Numerical implementation was performed in Python and is available at github.com/lukearcus/fossil\_scenario.}
% \textcolor{red}{We considered using a neural network...}
% Simulations were carried out on a server with 80 2.5 GHz CPUs and 125 GB of RAM. 
For all numerical simulations, we considered a confidence level of $\beta=10^{-5}$, $N=1000$ samples; our results are averaged across $5$ independent repetitions, each with different multi-samples. 
By sample complexity, we refer to the number of \emph{trajectory samples}, separate to the states used for the sample-independent loss since these samples can be obtained without accessing the system dynamics.

\subsection{Benchmark Dynamical System}
\label{sec:exp:main}

\begin{figure}[t]
\centering
	\includegraphics[width=0.75\linewidth]{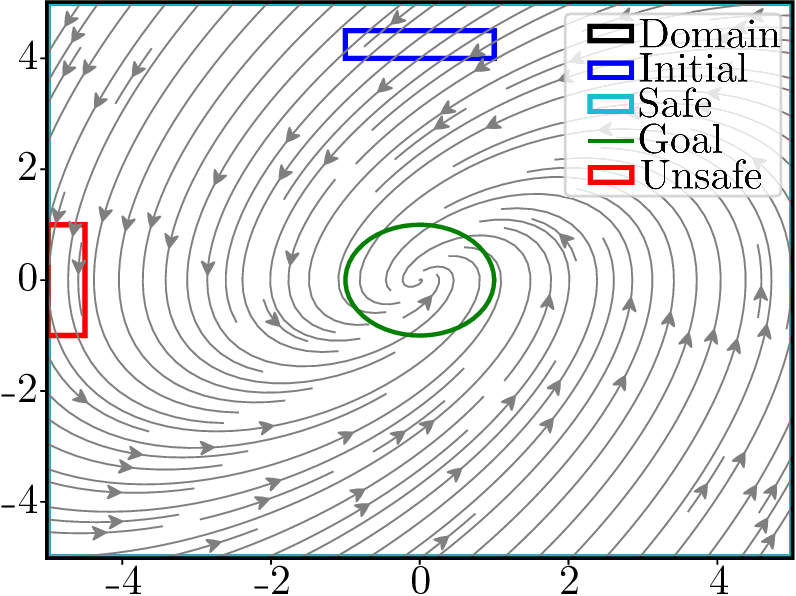}
	\caption{Phase plane plot for the dynamical system of \eqref{eq:spiral_dyn}. The different sets shown are related to the sets that appear in the definitions of the reachability, safety and RWA property. For each case, only the relevant sets are considered.}
	\label{fig:spiral_phase}
\end{figure}

To demonstrate the efficacy of our techniques across all certificates presented, we use the following dynamical system as benchmark, with state vector $x(k) = (x_1(k), x_2(k)) \in \mathbb{R}^2$, namely,
\begin{equation}
	\label{eq:spiral_dyn}
	\begin{bmatrix}x_1(k+1)\\x_2(k+1)\end{bmatrix}=\begin{bmatrix}x_1(k)-\frac{T_\text{d}}{2}x_2(k)\\ x_2(k)+\frac{T_\text{d}}{2}(x_1(k)-x_2(k))\end{bmatrix},
\end{equation}
where $T_\text{d} = 0.1$ and we use time horizon $T=100$ steps.
We used a neural network with 2 hidden layers, and 5 neurons per layer, with sigmoid activation functions thus leading to a parameter vector of size 51.
The phase plane plot for these dynamics is in Figure \ref{fig:spiral_phase} alongside different sets related to the definition of reachability, safety and RWA properties are shown. 
% For the reachability property we aim at verifying that trajectories reach the goal set $X_G$ (circle centered at the origin), without 
% leaving the domain $X$ until then. The unsafe set is not relevant as far as this property is concerned
% For safety, we require that trajectories do not enter the unsafe region $X_U$.
% Finally, for the RWA property, we have the domain $X$ and the unsafe set $X_U$.

\begin{figure*}[ht]
\centering
	\begin{minipage}{0.3\linewidth}
    \centering
		\includegraphics[width=\linewidth]{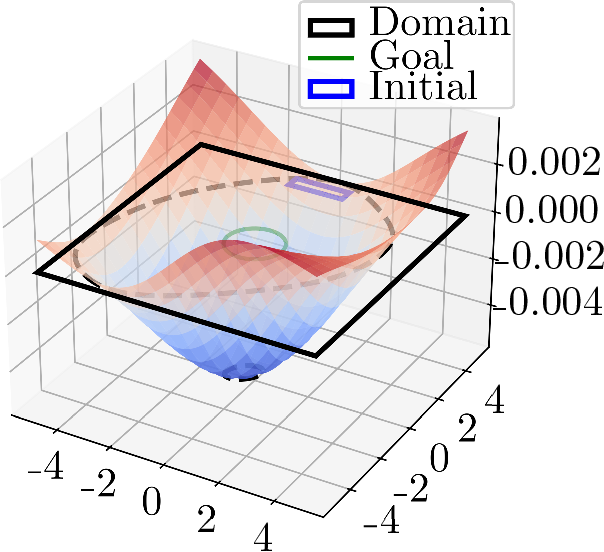}
        \caption{Surface plot of the reachability certificate}
        \label{fig:spiral_reach_surf}
	\end{minipage}\hfill
	\begin{minipage}{0.3\linewidth}
    \centering
		\includegraphics[width=\textwidth]{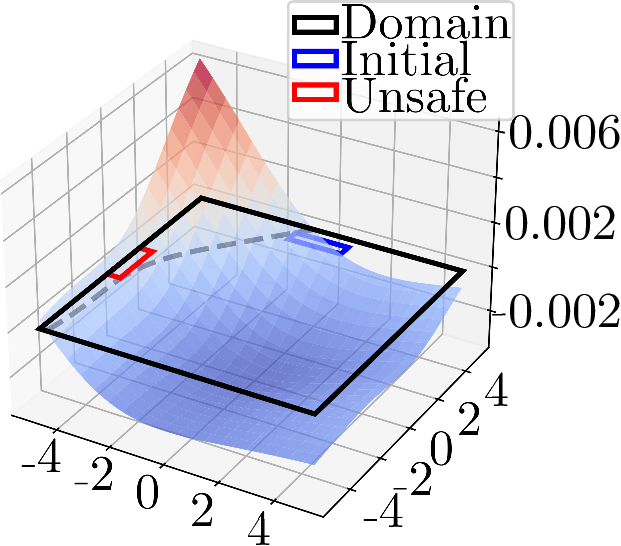}
        \caption{Surface plot of the safety/barrier certificate.}
        \label{fig:spiral_barr_surf}
	\end{minipage}\hfill
	\begin{minipage}{0.3\linewidth}
		\includegraphics[width=\linewidth]{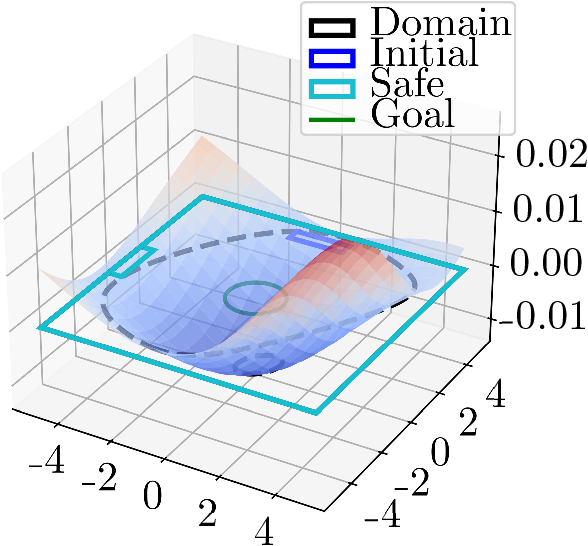}
        \caption{Surface plot of the RWA certificate.}
        \label{fig:spiral_rws_surf}
	\end{minipage}\hfill
\end{figure*}

\begin{table*}[t]
\begin{center}
    	\caption{Probabilistic guarantees for the system in \eqref{eq:spiral_dyn}. Standard deviations are shown in parentheses alongside means.}
	\begin{tabular}{p{2.5cm} || p{2cm} | p{2cm}| p{2cm} | p{2cm}| p{2cm}| p{2cm} }
		& Certificate Risk Bound $\varepsilon$ in Theorem \ref{thm:Guarantees} & Empirical Certificate Risk $\hat{\varepsilon}$&  Algorithm~\ref{algo:main} Computation Time (s)  & Property ~~~Risk Bound $\varepsilon$ in Prop. \ref{corr:a_post} & Empirical Property Risk $\hat{\varepsilon}$ & Direct Bound Computation Time (s) \\ \hline \hline
		Reach Certificate (Proposition \ref{cert:reach}) & 0.026 (0.004) & 0 (0) & 16597 (9157)  & 0.018 (0) & 0 (0) & 2.5 (0.3)  \\ \hline
		Safety Certificate (Proposition \ref{cert:barr}) & 0.052 (0.017) & 0 (0) & 7924 (505) & 0.018 (0)& 0 (0) & 7 (1) \\ \hline
		RWA Certificate ~~~(Proposition \ref{cert:RWA}) & 0.037 (0.021) & 0 (0) &  20120 (15783) & 0.018 (0) & 0 (0) & 7 (1)  \\ \hline 
	\end{tabular} \label{tab:guarantees}
\end{center}
\end{table*}

Surface plots of the reachability, barrier and RWA certificate are shown in Figure \ref{fig:spiral_reach_surf}, Figure \ref{fig:spiral_barr_surf} and Figure \ref{fig:spiral_rws_surf}, respectively. The zero and $-\delta$-sublevel sets of these certificates are highlighted with dashed black lines. With reference to Figure \ref{fig:spiral_reach_surf} notice that the zero-sublevel set includes both the initial and the goal set, and no states outside the domain as expected. Similarly, in Figure \ref{fig:spiral_barr_surf} the zero-sublevel set of the barrier function does not pass through the unsafe set, while the zero-sublevel set of the RWA certificate does not pass through the unsafe set, and does not include states outside the domain.

The constructed certificates depend on $N$ samples. By means of Algorithm \ref{algo:main} and Theorem \ref{thm:Guarantees}, these certificates are associated with a theoretical risk bound $\varepsilon$ (that bounds the probability that the certificate will not meet the conditions of the associated property when it comes to a new sample/trajectory). Table \ref{tab:guarantees} shows this risk bound as computed via Theorem \ref{thm:Guarantees}. 
We quantified empirically this property; namely, we generated additional samples and calculated the number of samples for which the computed certificate violated the associated certificate's conditions, or the underlying property. 
The number violating the certificate conditions (empirical certificate risk) is shown in the second column of Table \ref{tab:guarantees}, and the number violating the property (empirical property risk) is shown in the fifth column. 
Note that, as expected, the empirical values are lower than the theoretical bounds. 

The fourth column of Table \ref{tab:guarantees} provides the risk bound $\varepsilon$ that would be obtained for direct property violation statements (however, without allowing for certificate construction) as per Proposition \ref{corr:a_post}, this always results in a risk of 0.01825 as no samples are discarded, since the system can be shown to be deterministically safe. 
Recall that the results in the first column of Table \ref{tab:guarantees} bound (implicitly) the probability of property violation, as discussed in the second remark after the proof of Theorem \ref{thm:Guarantees}.

\subsection{Dynamical System of Higher Dimension}

We now investigate a dynamical system of higher dimension with a state $x(k) \in \mathbb{R}^8$, governed by
\begin{equation}
	\begin{aligned}
		x_i(k+1) &= x_i(k) + 0.1x_{i+1}(k),~ i = 1\dots 7,\\
		x_8(k+1) &= x_8(k)- 0.1(576x_1(k)+2400x_2(k)\\
        &+4180x_3(k)+3980x_4(k)+2273x_5(k)\\
             &+800x_6(k)+170x_7(k)+20x_8(k)).
	\end{aligned}
\end{equation}
We define $X=[-2.2,2.2]^8, X_I=[0.9,1.1]^8, X_U = [-2.2,-1.8]^8$ 
and use a neural network with 2 hidden layers, and 10 neurons per layer, with sigmoid activation functions thus leading to a parameter vector of size 211.
Once again, the entire of the initial set can be shown to be safe, so we aim to generate a guarantee as close to $0$ as possible.
We employ Algorithm \ref{algo:sub} to generate a safety certificate. This required an average of $0.273$ seconds, with a standard deviation of $0.018$ seconds. 

This certificate is computed much faster than those in Table~\ref{tab:guarantees}, which is possible since the runtime of our algorithm is primarily constrained by how many samples need to be removed by Algorithm~\ref{algo:main} in order to bring the loss to $0$.
This can be seen as a measure of how ``hard'' the problem is.
In this example, it is likely that the sets are easy to separate whilst still maintaining the difference condition, whereas the system in the previous section required more computation since trajectories move towards the unsafe set, before moving away from it. 

Due to the higher-dimensional state space, this certificate is not illustrated pictorially. 
It is accompanied by a probabilistic certificate $\varepsilon = 0.019$ (standard deviation $0.001$) computed by means of Theorem \ref{thm:Guarantees}.
Using Proposition~\ref{corr:a_post}, we find a guarantee of $0.018$ (standard deviation $0$), after $2.26$ seconds (standard deviation $0.05$s).

 \begin{figure*}[t]
 	\begin{minipage}[t]{0.3\linewidth}
	   \includegraphics[width=\linewidth]{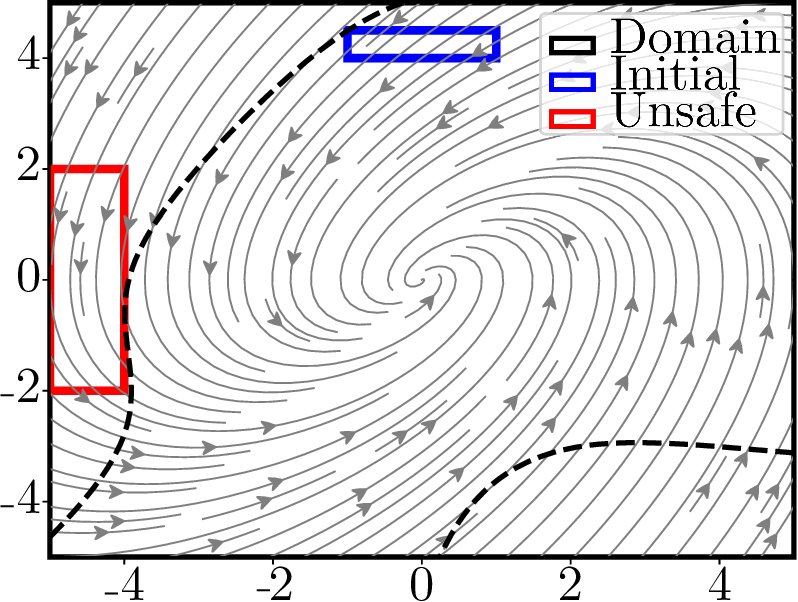}
	   \caption{Phase plane plot, initial and unsafe set for of partially unsafe system.}
	   \label{fig:spiral_barr_unsafe_plane} 
    \end{minipage}\hfill
 	\begin{minipage}[t]{0.3\linewidth}
	   \includegraphics[width=\linewidth]{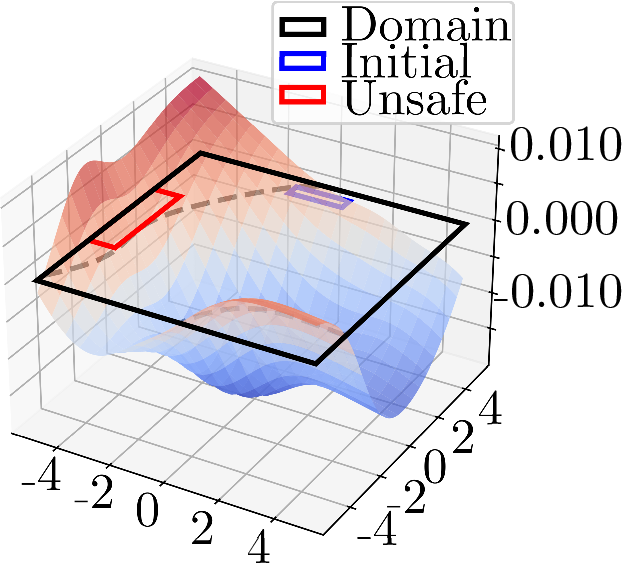}
	   \caption{Surface plot of the safety/barrier certificate for the partially unsafe system of Figure \ref{fig:spiral_barr_unsafe_plane}.}
	   \label{fig:spiral_barr_unsafe_surf}
    \end{minipage}\hfill
    \begin{minipage}[t]{0.3\linewidth}
        	\includegraphics[width=\linewidth]{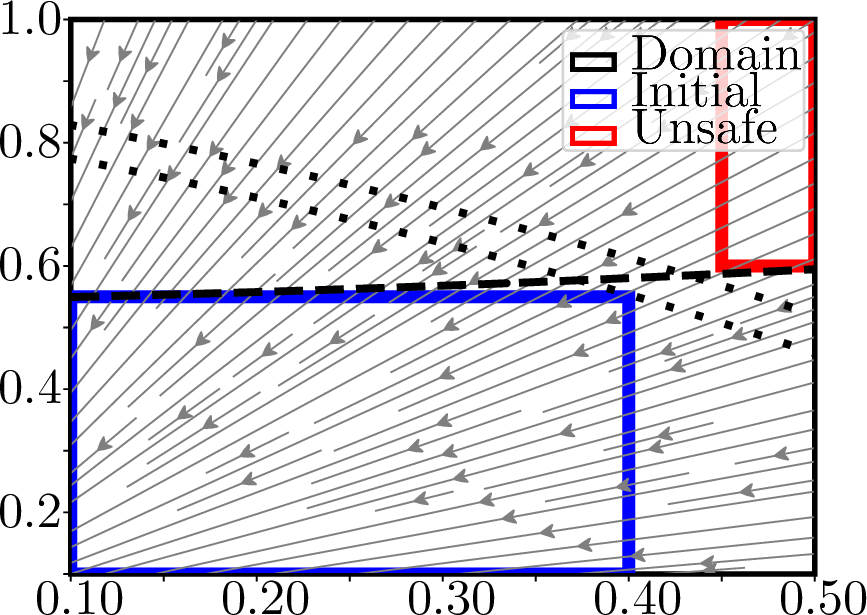}
	\caption{Comparison with \cite{DBLP:journals/tac/NejatiLJSZ23}. The zero-level set of the safety certificate of our approach is dashed; level sets that separate the initial and unsafe sets (i.e. $\gamma$- and $\lambda$- level sets) from \cite{DBLP:journals/tac/NejatiLJSZ23} are dotted.}
	\label{fig:jet_eng_comparison}
    \end{minipage}
 \end{figure*}
 
\subsection{Partially Unsafe Systems}
% For the numerical experiments so far, all sampled trajectories satisfied the property of interest.
We now consider the problem of safety certificate construction for the system in \eqref{eq:spiral_dyn} with an enlarged unsafe region (see Figure \ref{fig:spiral_barr_unsafe_plane}).
We employ the same neural network as in Section~\ref{sec:exp:main}.
We refer to this system as partially unsafe, as some sampled trajectories enter the unsafe set. Unlike existing techniques which require either a deterministically safe system~\cite{DBLP:conf/hybrid/EdwardsPA24}, or stochastic dynamics~\cite{DBLP:conf/cdc/PrajnaJP04}, we are still able to synthesize a probabilistic barrier certificate.  
The zero-sublevel set of the constructed safety certificate is shown by a dashed line in both Figures \ref{fig:spiral_barr_unsafe_plane} and \ref{fig:spiral_barr_unsafe_surf}. Figure \ref{fig:spiral_barr_unsafe_surf} provides a surface of the constructed certificate, and demonstrates that it separates the initial and the unsafe set. The computation time was $17971$ seconds (standard deviation $1414$s). 

For this certificate, we obtained a theoretical risk bound $\varepsilon = 0.388$ (standard deviation $0.035$) by means of Theorem \ref{thm:Guarantees}, and an empirical property risk of $\hat{\varepsilon} = 0.011$ (standard deviation $0.002$). 
% These guarantees are not tight; we could improve these by considering additional samples and performing the discarding procedure of Algorithm \ref{algo:main}, however, this would lead to larger computation times.
% To prevent discarding too many trajectories, we only discard those in the compression set at each iteration, which is likely to be a smaller number.
Proposition~\ref{corr:a_post} gives risk bound $0.042$ (standard deviation $0.004$) after $3.2$ seconds (standard deviation $0.1$s).

\subsection{Comparison with \cite{DBLP:journals/tac/NejatiLJSZ23}}

We extend the comparison of our work with that in \cite{DBLP:journals/tac/NejatiLJSZ23}, which has been reviewed in Section \ref{sec:related}. To this end, we construct a safety certificate for a two-dimensional DC Motor as considered in \cite{DBLP:journals/tac/NejatiLJSZ23}, using a neural network with 1 hidden layer, and 5 neurons per layer, with sigmoid activation functions thus leading to a parameter vector of size 21. 
We first replicate the methodology of \cite{DBLP:journals/tac/NejatiLJSZ23}, using the Lipschitz constants they provide.

The methodology of \cite{DBLP:journals/tac/NejatiLJSZ23} required $257149$ samples and $307$ seconds (standard deviation $44$s) of computation time to compute a barrier certificate with confidence at least equal to $0.99$. 
Using $1000$ samples and $1.4$ seconds of computation time (standard deviation $0.1$s), we obtained $\varepsilon = 0.01$ (standard deviation $0$), i.e. we can bound safety with a risk of 1\%, for the same confidence.
It can thus be observed that the numerical computation savings (in terms of number of samples -- this might be an expensive task -- and computation time) are significant. Figure \ref{fig:jet_eng_comparison} illustrates a phase plane plot and the initial and unsafe sets for this problem. The dotted lines correspond to the sublevel sets constructed in \cite{DBLP:journals/tac/NejatiLJSZ23} (one lower bounding the unsafe set, the other upper bounding the initial set). 
The dashed line depicts the zero-sublevel set of the certificate constructed by our approach.

We also performed a comparison on the following four-dimensional system, a discretized version of a model taken from~\cite{DBLP:conf/hybrid/EdwardsPA24}, with the required Lipschitz constants estimated using the technique in \cite{DBLP:journals/jgo/WoodZ96}.
% \begin{equation}
\begin{align}
        x_1(k+1) &= x_1(k) + 0.1\left(\frac{x_1(k)  x_2(k)}{5} -\frac{x_3(k)x_4(k)}{2}\right),\nonumber\\
        x_2(k+1) &= x_2(k)+0.1\cos(x_4(k)),\\
        x_3(k+1) &= x_3(k)+0.001\sqrt{|x_1(k)|},\nonumber\\
        x_4(k+1) &= x_4(k)+0.1\left(-x_1(k) - x_2(k)^2 + \sin(x_4(k))\right)\nonumber.
\end{align}
% \end{equation}
Due to the reasons outlined in Section \ref{sec:related}, the approach of \cite{DBLP:journals/tac/NejatiLJSZ23} with $10^{19}$ samples results in a confidence of at least $10^{-30}$, which is not practically useful.
In contrast, with our techniques with $1000$ samples we obtain a risk level of $\varepsilon = 0.02039$, with confidence at least $1-10^{-5}$.

\section{Conclusions}
\label{sec:conc}

We have proposed a method for synthesis of neural-network certificates, based only on a finite number of trajectories from a system, in order to verify a number of core temporally extended specifications.  
% We have proposed a method for synthesis of neural-network certificates, based only on a finite number of trajectories from a system, in order to verify a number of core temporally extended specifications.  
These certificates allow providing assertions on the satisfaction of the properties of interest. 
In order to synthesize a certificate, we considered a novel algorithm for solving a non-convex optimization program where the loss function we seek to minimize encodes different conditions on the certificate to be learned. 

As a byproduct of our algorithm, we determine a quantity termed ``compression set'', which is instrumental in obtaining scalable probabilistic guarantees. 
% This process is novel per se and provides a constructive mechanism for compression set calculation, thus opening the road for its use to more general
% non-convex optimization problems.
Our numerical experiments demonstrate the efficacy of our methods on a number of examples, involving comparison with related methodologies in the literature. 

% Current work concentrates towards extending our analysis to continuous-time dynamical systems, as well as on considering controlled systems, thus co-designing a controller and a certificate at the same time.

\bibliographystyle{myplain}
\bibliography{neural_certs_paper}

\ifappendix
\appendix   
\section{Proofs}
\label{app:proofs}

\subsection{Certificate Proofs}

\subsubsection{Proof of Proposition \ref{cert:reach} -- Reachability Certificate}
Fix $\delta > -\sup_{x \in X_I} V(x) \geq 0$, and recall that $k_G = \min \{k \in \{0,\dots,T\} \colon V(x(k)) \leq -\delta\}$. Consider then the difference condition in \eqref{eq:reach_deriv}, namely,
    \begin{equation} \label{eq:proof_dec_reach}
        \begin{aligned}
		&V(x(k+1)) - V(x(k)) \\ 
        & < - \frac{1}{T}\Big (\sup_{x \in X_I} V(x)+\delta \Big),~ k=0,\dots,k_G-1,
        \end{aligned}
    \end{equation}
By recursive application of this inequality $k \leq k_G$ times, 
\begin{align}
&V(x(k)) < V(x(0)) -\frac{k}{T} \Big ( \sup_{x \in X_I} V(x) +\delta\Big ) \nonumber \\
&\leq \frac{T-k}{T} \sup_{x \in X_I} V(x) -\frac{k}{T} \delta  \leq -\frac{k}{T} \delta \leq 0, \label{eq:proof_dec_reach1}
\end{align}
where the second inequality is since $V(x(0)) \leq \sup_{x \in X_I} V(x)$, as $x(0) \in X_I$. The third
one is since $\sup_{x \in X_I} V(x) \leq 0$ as by \eqref{eq:reach_init}, $V(x) \leq 0$, for all $x \in X_I$, and  $k \leq k_G \leq T$, while the last inequality is since $\delta>0$.

By \eqref{eq:proof_dec_reach1} we then have that for all $k\leq k_G$, $V(x(k)) <0$, which implies that $x(k)$ does not leave $X$ for all $k\leq k_G$ (see \eqref{eq:reach_else}), while by the definition of $k_G$, $x(k_G) \in X_G$. Notice that if $k_G = T$, then \eqref{eq:proof_dec_reach1} (besides implying that $x(k) \in X$ for all $k\leq T$), also leads to $V(x(T)) \leq -\delta$, which means that $x(T) \in X_G$ after $T$ time steps (see \eqref{eq:reach_goal}), which captures the latest time the goal set is reached.

Therefore, all trajectories that start within $X_I$ reach the goal set $X_G$ in at most $T$ steps, without escaping $X$ till then, thus concluding the proof. \qed

\subsubsection{Proof of Proposition \ref{cert:barr} -- Safety Certificate}

Consider the condition in \eqref{eq:barr_deriv}, namely,
    \begin{equation}
     \begin{aligned}
        &V(x(k+1))-V(x(k)) \\ &< \frac{1}{T} \Big( \inf_{x \in X_U}V(x)-\sup_{x \in X_I}V(x) \Big ),~ k=0,\dots,T-1.
     \end{aligned}
     \end{equation}
     By recursive application of this inequality for $k\leq T$ times, we obtain 
     \begin{align}
         &V(x(k)) < V(x(0)) + \frac{k}{T} \Big( \inf_{x \in X_U}V(x)-\sup_{x \in X_I}V(x) \Big ) \nonumber \\
         &\leq \frac{T-k}{T} \sup_{x \in X_I}V(x) + \frac{k}{T} \inf_{x \in X_U}V(x) \nonumber \\
         &\leq \frac{k}{T} \inf_{x \in X_U}V(x) \leq \inf_{x \in X_U}V(x). \label{eq:proof_safety}
     \end{align}
     where the second inequality is is since $V(x(0)) \leq \sup_{x \in X_I}V(x)$, as $x(0) \in X_I$. The third inequality is since $\sup_{x \in X_I}V(x) \leq 0$ as by \eqref{eq:barr_states1}, $V(x) \leq 0$ for all $x \in X_I$ and $k\leq T$. The last inequality is since $\inf_{x \in X_U}V(x)\geq 0$, as by \eqref{eq:barr_states2} $V(x) >0$ for all $x \in X_U$, and $k\leq T$.
     We thus have
           \begin{align}
       V(x(k)) &< \inf_{x \in X_U}V(x),~ k=1,\dots,T.
       \end{align}
       and hence $x(k) \notin X_U, k=0,\dots,T$ (notice that $x(0) \notin X_U$ holds since $X_I \cap X_U = \emptyset$). The latter implies that all trajectories that start in $X_I$ avoid entering the unsafe set $X_U$, thus concluding the proof. \qed

\subsubsection{Proof of Proposition \ref{cert:RWA} -- RWA Certificaate}

% Fix $\delta > -\sup_{x \in X_I} V(x) \geq 0 $, and recall that $k_G = \min \{k \in \{0,\dots,T\} \colon V(x(k)) \leq -\delta\}$.
% Consider then the difference condition in \eqref{eq:RWA_deriv1}, namely,
%     \begin{equation} \label{eq:proof_RWA}
%      \begin{aligned}
%         &V(x(k+1))-V(x(k)) \\ &<-\frac{1}{T} \Big ( \sup_{x \in X_I}V(x)+\delta \Big ),~ k=0,\dots,k_G-1,
%      \end{aligned}
%      \end{equation}
%     Note that this is identical to the difference condition for our reachability property, and hence following the same arguments with the proof of Proposition \ref{cert:reach}, we can infer that state trajectories emanating from $X_I$ will reach the goal set $X_G$ in at most $T$ time steps.
    Since we must satisy $\psi_\mathrm{reach}$, we can cinclude that, following Proposition \ref{cert:reach}, state trajectories emanating from $X_I$ will reach the goal set $X_G$ in at most $T$ time steps.
    
    By \eqref{eq:barr_states2} we have that $V(x) > 0$, for all $x \in _U$ while by \eqref{eq:reach_init} we have that $V(x) \leq 0$, for all $x \in X_I$. Therefore, $\sup_{x\in X_I} V(x) \leq 0 \leq \inf_{x \in X_U} V(x)$. 
    At the same time by our choice for $\delta$ we have that $\delta>-\sup_{x \in X_I}V(x)$. Combining these, we infer that $\delta >-\inf_{x \in X_U} V(x)$.
    Thus, \eqref{eq:reach_deriv} implies that for all $k=0,\ldots,k_G-1$,
    \begin{align}
        -\frac{1}{T} \Big ( \sup_{x \in X_I}&V(x) +\delta \Big )  \nonumber \\
        &< \frac{1}{T} \Big (\inf_{x \in X_U}V(x)-\sup_{x \in X_I}V(x)\Big ).
    \end{align}
    Therefore,
     \begin{align}
        &V(x(k+1))-V(x(k)) \\ &<\frac{1}{T} \Big (\inf_{x \in X_U}V(x)-\sup_{x \in X_I}V(x)\Big ), k=0,\dots,k_G-1. \nonumber
     \end{align}
     Note that this is identical to the difference condition for our safety property, and hence following the same arguments with the proof of Proposition \ref{cert:barr}, we can infer that state trajectories emanating from $X_I$ will never pass through the unsafe set $X_U$ until time $k=k_G$.

     Moreover, by \eqref{eq:RWA_deriv2}, we have that 
      \begin{equation}
     \begin{aligned}
        &V(x(k+1))-V(x(k)) \\ &<\frac{1}{T} \Big (\inf_{x \in X_U}V(x)+\delta\Big ),~ k=k_G,\dots,T-1.
     \end{aligned}
     \end{equation}
     Note that this is also a difference condition identical to that for our safety property, but with $\delta$ in place of $\sup_{x \in X_I} V(x)$ (since we know that $V(x(k_G)) \leq -\delta$ by definition of $k_G$).
     Hence, we have a safety condition for all trajectories emanating from this sublevel set.
     We know that trajectories reach this sublevel set, and hence remain safe for $k=k_G,\ldots,T$.

    Therefore, we have shown that starting at $X_I$ trajectories reach $X_G$ in at most $T$ time steps, while they never pass through $X_U$, thus concluding the proof. \qed

\subsection{Proof of Proposition \ref{prop:converge} -- Properties of Algorithm 1}
\begin{enumerate}[wide, labelwidth=!, labelindent=0pt]
\item By construction, Algorithm \ref{algo:sub} creates a non-increasing sequence of iterates $\{L_k\}_{k\geq 0}$ that is bounded below by the global minimum of $\min_{\xi \in \mathcal{D}} L(\theta,\xi)$ which exists and is finite due to Assumption \ref{ass:exist}.
As such, the sequence $\{L_k\}_{k\geq 0}$ is convergent, which in turn implies that Algorithm \ref{algo:sub} terminates.\\
\item We need to show that the set $\mathcal{C}_N$ is a compression set in the sense of Definition \ref{def:compress} with $\mathcal{A}$ being Algorithm \ref{algo:sub} with $\mathcal{D} = \{\xi_i\}_{i=1}^N$. 
To see this, we ``re-run'' Algorithm \ref{algo:sub} from the same initial choice of the parameter vector $\theta$ but with $\mathcal{C}_N$ in place of $\mathcal{D}$. 
Notice that exactly the same iterates will be generated, as $\mathcal{C}_N$ contains all samples that have a misaligned subgradient and value greater than the loss evaluated on the running compression set. %all samples that have led to a worst case loss across iterations (step 6). 
As a result, the same output will be returned, which by Definition \ref{def:compress} establishes that $\mathcal{C}_N$ is a compression set.\\
\item We show that all properties of Assumption \ref{ass:alg_prop} are satisfied by Algorithm \ref{algo:sub}. \\

\emph{Preference:} Consider a fixed (sample independent) initialization of Algorithm \ref{algo:sub} in terms of the parameter $\theta$. 
Consider also any subsets $\mathcal{C}_1,\mathcal{C}_2$ of 
$\{\xi^i\}_{i=1}^N$ with $\mathcal{C}_1\subseteq \mathcal{C}_2$. 

Suppose that the compression set returned by Algorithm \ref{algo:sub} when fed with $\mathcal{C}_2$ is different from 
$\mathcal{C}_1$. 
          Fix any $\xi \in \Xi$ and consider the set $\mathcal{C}_2 \cup \{\xi\}$. We will show that the compression set returned by Algorithm \ref{algo:sub} when fed with $\mathcal{C}_2 \cup \{\xi\}$ is different from $\mathcal{C}_1$ as well.\\
          \emph{Case 1}: The new sample $\xi$ does not appear as a maximizing sample in step~\ref{line:max_D} of Algorithm \ref{algo:sub}, or its subgradient is such that the quantity in step~\ref{line:inner} is positive. This implies that step~\ref{line:update_C} is not performed and the algorithm proceeds directly to step~\ref{line:approx_step}. As such, $\xi$ is not added to the compression set returned by Algorithm \ref{algo:sub}, which remains the same with that returned when the algorithm is fed only by $\{\xi^i\}_{i=1}^N$. However, the latter is not equal to $\mathcal{C}_1$, thus establishing the claim.\\  
	       \emph{Case 2}: The new sample $\xi$ appears as a maximizing sample in step~\ref{line:max_D} of Algorithm \ref{algo:sub}, and has a subgradient such that the quantity in step~\ref{line:inner} is non-positive. As such, step~\ref{line:update_C} is performed and $\xi$ is added to the compression returned by Algorithm \ref{algo:sub}.
          If $\xi \notin \mathcal{C}_1$ then the resulting compression set will be different from $\mathcal{C}_1$ as it would contain at least one element that is not $\mathcal{C}_1$, namely, $\xi$.
          
	       If $\xi \in \mathcal{C}_1$ then it must also be in $\mathcal{C}_2$ as $\mathcal{C}_1 \subseteq \mathcal{C}_2$. In that case $\xi$ would appear twice in $\mathcal{C}_2 \cup \{\xi\}$, i.e., the set of samples with which Algorithm \ref{algo:sub} is fed has $\xi$ as a repeated sample (notice that this can happen with zero probability due to Assumption \ref{ass:non-conc_mass}). 
                           
          Once one of these repeated samples is added to the compression set returned by Algorithm \ref{algo:sub}, then the other will never be added. This is since when this other sample appears as a maximizing one in step~\ref{line:max_D} then its duplicate will already be in the compression set, and hence the exact and approximate subgradients in steps~\ref{line:subgrad_D} and~\ref{line:subgrad_C} would be identical. 
          As such, the quantity in step~\ref{line:inner} would be non-negative (and, by positive-definiteness of the inner product, only zero when both vectors are zero-vectors) and hence step \ref{line:update_C} will not be performed, with the duplicate not added to the compression set. 
          As such, one of the repeated $\xi$'s is redundant, which implies that the compression set returned by Algorithm \ref{algo:sub} when fed with $\mathcal{C}_2 \cup \{\xi\}$ is the same with the one that would be returned when it is fed with $\mathcal{C}_2$. 
          However, this would imply that if $\mathcal{C}_1$ is the compression returned by Algorithm \ref{algo:sub} when fed with of $\mathcal{C}_2 \cup \{\xi\}$, it will also be the compression set for $\mathcal{C}_2$ (as the duplicate $\xi$ would be redundant). 
          However, the starting hypothesis has been that $\mathcal{C}_1$ is not a compression of $\mathcal{C}_2$. As such, it is not possible for $\mathcal{C}_1$ to be a compression set of $\mathcal{C}_2 \cup \{\xi\}$ as well, establishing the claim.\\
          
\emph{Non-associativity:} Consider a fixed (sample independent) initialization of Algorithm \ref{algo:sub} in terms of the parameter $\theta$. Let $\{\xi^i\}_{i=1}^{N+\bar{N}}$ for some $\bar{N} \geq 1$.
        Suppose that $\mathcal{C}$ is returned by Algorithm \ref{algo:sub} a compression set of $\{\xi^i\}_{i=1}^{N} \cup \{\xi\}$, for all $\xi \in \{\xi^i\}_{i=N+1}^{N+\bar{N}}$. 
        Therefore, up to a measure zero set we must have that
        \begin{align}
        \mathcal{C} \subset \bigcap_{j=N+1}^{\bar{N}}
       \Big ( \{\xi^i\}_{i=1}^N \cup \{\xi^j\} \Big ) = \{\xi^i\}_{i=1}^N, \label{eq:proof_nonassoc}
        \end{align}
where the inclusion is since $\mathcal{C}$ is assumed to be returned as a compression set by Algorithm \ref{algo:sub} when this is fed with any set within the intersection, while the equality is since by Assumption \ref{ass:non-conc_mass} all samples in $\{\xi^i\}_{i=1}^{N+\bar{N}}$ are distinct up to a measure zero set. This implies that up to a measure zero set $\mathcal{C}$ should be a compression set returned by Algorithm \ref{algo:sub} whenever this is fed with $\{\xi^i\}_{i=1}^N$ as any additional sample would be redundant.

Fix now any $\xi \in \{\xi^i\}_{i=N+1}^{N+\bar{N}}$, and consider 
Algorithm \ref{algo:sub} with $\mathcal{D} = \{\xi^i\}_{i=1}^{N} \cup \{\xi\}$. The fact that $\mathcal{C}$ is returned as a compression set for $\{\xi^i\}_{i=1}^{N} \cup \{\xi\}$ implies that whenever $\xi$ is a maximizing sample in step~\ref{line:max_D} of Algorithm \ref{algo:sub}, it should give rise to a subgradient such that the quantity in step $10$ of the algorithm is positive. This implies that step~\ref{line:approx_step} is performed and hence $\xi$ is not added to $\mathcal{C}$. 

Considering Algorithm \ref{algo:sub} this time with $\mathcal{D} =\{\xi^i\}_{i=1}^{N+\bar{N}}$, i.e., fed with all samples at once, due to the aforementioned arguments, whenever a $\xi \in \{\xi^i\}_{i=N+1}^{N+\bar{N}}$ is a maximizing sample in step~\ref{line:max_D}, then the algorithm would proceed to step~\ref{line:approx_step}, and steps~\ref{line:true_step}--\ref{line:update_C} will not be executed. As such, no such $\xi$ will be added to $\mathcal{C}$. 

Hence, the compression set returned by Algorithm \ref{algo:sub} when fed with $\{\xi^i\}_{i=1}^{N+\bar{N}}$ would be the same with the one that would be returned if the algorithm was fed with $\{\xi^i\}_{i=1}^{N}$. By \eqref{eq:proof_nonassoc} this then implies that the returned set should be $\mathcal{C}$ up to a measure zero set. \qed
\end{enumerate}

\subsection{Proof of Proposition~\ref{prop:converge_main}}

\begin{enumerate}[wide, labelwidth=!, labelindent=0pt]
\item At every iteration, Algorithm~\ref{algo:sub}, is called with fewer samples, and initialized on the optimal parameter set from the previous iteration. 
Hence, the loss value is a non-increasing sequence.
If all samples are removed, the loss is zero, since we optimize only the sample-independent loss.
Hence, the sequence converges to zero (in the worst-case upon removing all samples).
\item 
Consider Algorithm \ref{algo:main} with $\mathcal{D} = \{\xi_i\}_{i=1}^N$.
Denote by $\mathcal{C}_i$ the set returned at step $5$ of Algorithm \ref{algo:main}, and recall that $\mathcal{C}_i \subseteq \mathcal{D}$ is the compression set returned by Algorithm \ref{algo:sub} when this is invoked at that part of the process. Notice then that the set $\mathcal{R}_N$ returned by Algorithm \ref{algo:main} can be expressed as $\mathcal{R}_N = \bigcup_{i}\mathcal{C}_i$.\\
We need to show that $\mathcal{R}_N$ is a compression set in the sense of Definition \ref{def:compress} with $\mathcal{A}$ being Algorithm \ref{algo:main} with $\mathcal{D} = \{\xi_i\}_{i=1}^N$. 
To see this, we ``re-run'' Algorithm \ref{algo:main} from the same initial choice of the parameter vector $\theta$ but with $\mathcal{R}_N$ in place of $\mathcal{D}$. At the first iteration, the set returned in step~\ref{line:subgrad} is $\mathcal{C}_1$ (and the parameter returned would be the same with the one that would be obtained if all samples were employed) as this is a compression set for Algorithm \ref{algo:sub} invoked at that step with $\mathcal{D} = \mathcal{R_N}$. As such, in step~\ref{line:discard} and~\ref{line:update_outer_C} we would, respectively, have that $\mathcal{D} = \mathcal{R}_N \setminus \mathcal{C}_1$, and $\mathcal{R} = \mathcal{R}_N$ since $\mathcal{C}_1$ is already in $\mathcal{R}_N$. Proceeding analogously, we have that Algorithm \ref{algo:main} terminates with the set $\mathcal{R}$ remaining intact to $\mathcal{R}_N$ and $\mathcal{D}$ being empty, and $\mathcal{R} = \mathcal{R}_N$. This establishes that $\mathcal{R}_N$ is a compression set for Algorithm \ref{algo:main}.
\item Since Algorithm~\ref{algo:sub} satisfies Assumption~\ref{ass:alg_prop}, and we simply call this algorithm repeatedly, then Algorithm~\ref{algo:main}, also inherits these properties and satisfies Assumption~\ref{ass:alg_prop}.
\end{enumerate}

\fi

\end{document}